\documentclass[aps,prd,nofootinbib,reprint,superscriptaddress,longbibliography]{revtex4-1}
\usepackage{etoolbox}
\usepackage{dcolumn}% Align table columns on decimal point
\usepackage{bm}% bold math
\usepackage{graphics}
\usepackage{dirtytalk}
\usepackage{graphicx} %Include figure filesusepackage{graphicx} %Include figure files
\usepackage{blindtext}
\usepackage{amsmath}
\usepackage{amsfonts}
\usepackage{amssymb}
\usepackage[%
colorlinks=true,
urlcolor=blue,
linkcolor=red,
citecolor=blue
]{hyperref}

\usepackage{wrapfig}
\usepackage{natbib}
\begin{document}
\title{\bf  Mutual information of subsystems and the Page curve for Schwarzschild de-Sitter black hole}
\vskip 1cm
	\vskip 1cm
\author{Anirban Roy Chowdhury}
\email{iamanirban.rkmvc@gmail.com}
\affiliation{Department of Astrophysics and High Energy Physics,\linebreak
	S.N.~Bose National Centre for Basic Sciences, JD Block, Sector-III, Salt Lake, Kolkata 700106, India}	
\author{Ashis Saha}
\email{ashisphys18@klyuniv.ac.in}
\affiliation{Department of Physics, University of Kalyani, Kalyani 741235, India}
\author{Sunandan Gangopadhyay}
\email{sunandan.gangopadhyay@bose.res.in}
\affiliation{Department of Astrophysics and High Energy Physics,\linebreak
	S.N.~Bose National Centre for Basic Sciences, JD Block, Sector-III, Salt Lake, Kolkata 700106, India}
%\date{}
	\begin{abstract}
	\noindent In this work, we show that the two proposals associated to the mutual information of matter fields can be given for an eternal Schwarzschild black hole in de-Sitter spacetime. These proposals also depicts the status of associated entanglement wedges and their roleplay in obtaining the correct Page curve of radiation. The first proposal has been give for the before Page time scenario, which shows that the mutual information $I(R_{H}^{+}:R_{H}^{-})$ vanishes at a certain value of the observer's time $t_{b_{H}}=t_{H}$ (where $t_{H}\ll \beta_{H}$). We claim that this is the Hartman-Maldacena time at which the entanglement wedge associated to $R_{H}^{+}\cup R_{H}^{-}$ gets disconnected and the fine-grained radiation entropy has the form $S(R_{H})\sim \log(\beta_{H})$. The second proposal depicts the fact that just after the Page time, when the replica wormholes are the dominating saddle-points, the mutual information $I(B_{H}^{+}:B_{H}^{-})$ vanishes as soon as the time difference $t_{a_{H}}-t_{b_{H}}$ equals the scrambling time. Holographically, this reflects that the entanglement wedge associated to $B_{H}^{+}\cup B_{H}^{-}$ jumps to the disconnected phase at this particular time-scale. Furthermore, these two proposals lead us to the correct time-evolution of the fine-grained entropy of radiation as portrayed by the Page curve. We have also shown that similar observations can be obtained for the radiation associated to the cosmological horizon.
	\end{abstract}
\maketitle
\noindent 
Hawking radiation is one of the most fascinating and mysterious phenomena in theoretical physics, and it is caused by pair formation that takes place in the black hole's near-horizon area \cite{Hawking:1975vcx}. This phenomenon has drawn a lot of attention in the context of modern theoretical physics. This is because, as a quantum mechanical radiation, its presence provides a clear indication of the microscopic physics underlying the general relativity theory. This has motivated to probe its quantum mechanical components, such von Neumann entropy \cite{Chuang:2000}. However, the investigation of the von Neumann entropy of the Hawking radiation has in turn provided us with a paradox. The paradox can be described in the following way. It has been noted that the creation of a black hole, which results from the gravitational collapse of a massive shell, is associated to a pure state. This implies that the corresponding von Neumann entropy is zero. Additionally, according to the theory of unitary evolution, the final state at the end of the evaporation process must likewise be a pure state, meaning that the von Neumann entropy once again must vanish at the end of the evaporation process. Hawking's semi-classical analysis, however, demonstrated that for an evaporating black hole, the von Neumann entropy of Hawking radiation is an ever-increasing quantity with regard to the observer's time \cite{PhysRevD.14.2460}, and it does not disappear even if the black hole has completely evaporated.\\
There is another way to understand this current scenario which is more suitable for the case of an eternal black hole. The von Neumann entropy of radiation is an well-known example of fine-grained type of entropy\footnote{It is also to be noted that the entanglement entropy of the radiation is identified as the von Neumann entropy of matter fields located on the region $R$ outside the black hole.} and on the other other hand, the thermodynamic entropy of the black hole is a perfect example of coarse-grained type of entropy \cite{Bekenstein:1972tm,Bekenstein:1973ur,PhysRevD.13.191}). Further, as the state corresponding to the whole system (radiation subsystem $R$ + black hole subsystem $R^c$) is a pure state, the fine-grained entropy of radiation is equal to the fine-grained entropy of the black hole subsystem, that is $S_{vN}(R)=S_{vN}(R^c)$. This observation together with the Hawking's semi-classical analysis implies that after a certain amount of time, the fine-grained entropy of black hole subsystem will greater than the coarse-grained entropy of the black hole ($S_{vN}(R)> S_{BH}$). This fact is self-contradictory as the basic definition of coarse-grained entropy is associated to the fact that it is obtained by maximizing the fine-grained entropy over all possible states. The above mentioned observation provides us an entropic way to understand the paradoxical situation.\\
So a natural question arises regarding the correct time evolution of the von Neumann entropy of Hawking radiation. This was efficiently addressed by the so-called \textit{Page curve}. The Page curve curve suggested that in order to satisfy the unitarity condition, the von Neumann entropy of the radiation shall start from zero and monotonically increase upto the \textit{Page time} and then again drop down to zero, signifying the end of the evaporation process \cite{PhysRevLett.71.3743,Page_2013}. The contradiction only emerges after the Page time as after this particular time one usually gets $S_{vN}(R)> S_{BH}$. Numerous intriguing methods have been developed to handle this problem while taking into account the unitarity evolution of radiation \cite{Almheiri:2012rt,Almheiri:2013hfa,Lloyd:2013bza,Papadodimas:2013wnh}. Recently, the idea of entanglement wedge reconstruction from Hawking radiation has proposed that certain regions in the interior of a black hole may be responsible for the fine-grained entropy of that radiation \cite{Penington:2019npb,Almheiri:2019psf,Almheiri:2019hni,Almheiri:2019yqk}. These auxiliary areas are known as \textit {islands}, and the surfaces at their ends are known as \textit {quantum extremal surfaces} (QES) \cite{Engelhardt:2014gca,Engelhardt:2019hmr,Akers:2019lzs,Wall:2012uf}. It is to be mentioned that the quantum extremal surfaces are the quantum corrected classical extremal surfaces \cite{PhysRevLett.96.181602,Hubeny:2007xt}. The fine-grained entropy of the Hawking radiation in the presence of the island in the black hole interior is provided by
\begin{equation}\label{eq1}
	S(R) =	\textrm{min}~\mathop{\textrm{ext}}_{\mathcal{\mathrm{I}}}\bigg\{\frac{\textrm{Area}(\partial I)}{4G_N}+S_{vN}(I\cup R)\bigg\}~.
\end{equation}
From a semi-classical perspective, the islands come from the replica wormhole saddle points (with the appropriate boundary conditions) of the gravitational path integral, which occur as a result of the use of the replica method in dynamical gravitational background \cite{Almheiri:2019qdq,Penington:2019kki,Goto:2020wnk,Colin-Ellerin:2020mva}. Due to this remarkable observation, the island formulation has emerged as an important prescription to be studied \cite{Almheiri:2020cfm,chen2020information,Hashimoto:2020cas,Hartman:2020swn,Anegawa:2020ezn,Dong:2020uxp,Balasubramanian:2020xqf,Raju:2020smc,Alishahiha:2020qza,Krishnan:2020fer,Krishnan:2020oun,Azarnia:2021uch,Yu:2021cgi,Arefeva:2021kfx,He:2021mst,Omidi:2021opl,Yu:2021rfg,Ahn:2021chg,Yadav:2022fmo,Du:2022vvg,HosseiniMansoori:2022hok,Ageev:2022qxv,Yu:2022xlh,Hu:2022zgy}.\\
It is important to note that while the majority of the above mentioned studies are restricted to the black holes in asymptotically flat or AdS spacetimes, the most recent finding indicates that our universe is of de-Sitter nature. Therefore, it makes sense to investigate the effect of the positive cosmological constant in context of the information paradox problem. Keeping this in mind, we will consider the eternal Schwarzschild de-Sitter (SdS) spacetime as the black hole spacetime in this paper. Given that these black holes are formed during the early inflationary stage of our universe, the information paradox problem for the Schwarzschild de-Sitter black holes is crucial. It also offers an ideal toy model for global structures of isolated black holes in our universe, keeping in mind the current phase of our universe's accelerated expansion. There are also causally disconnected areas in de-Sitter space, which is similar to the situation with black holes. Therefore, an observer may only access the regions of the universe that are enclosed by their own horizon. Furthermore, the cosmological event horizons emit and take in radiation similar to the black hole (Gibbons-Hawking radiation).
In general, the entropy creation of the cosmological horizon is an observer-dependent feature in contrast to the black hole. It is caused by a lack of knowledge about what exists outside of the cosmic horizon. In this work, we will try to obtain the correct Page curve for the black hole horizon of the SdS black hole and the Page-like curve for the cosmological horizon of the same black hole. We shall do this by keeping in mind the island formulation. It is also to be mentioned that apart from the approach (gravitational set up) which we have followed in this work, there is another way (gravitational set up) to address this entropic paradox. This is known as the doubly holographic set up \cite{takayanagi2011holographic,fujita2011aspects,nozaki2012central,PhysRevD.96.046005,Geng:2020qvw,Geng:2021wcq,Geng:2021iyq,hu2022ads,Geng:2021mic}. 
Some very interesting works in this set up can be found in \cite{Almheiri:2019psf,chen2020information,rozali2020information,almheiri2020entanglement,bousso2020gravity,Afrasiar:2023jrj,Afrasiar:2022ebi,Guo:2023gfa,Miao:2023unv,Hu:2022zgy}.\\
In \cite{Saha:2021ohr,RoyChowdhury:2022awr} it was shown that the mutual information of various subsystems plays a crucial role in obtaining the correct Page curve of Hawking radiation. To be precise, in \cite{Saha:2021ohr} it was shown that just after the Page time the mutual information of matter fields localized on $R_{+}$ and $R_{-}$ intervals vanishes, which eventually leads to a time-independent profile of fine-grained entropy $S(R)$.
Furthermore, in \cite{RoyChowdhury:2022awr} the previous observation was exploited in detail and two proposals were given regarding the saturation of mutual information (of various subsystems) for two different time domains (before and after the Page time). However, these works were only restricted to the eternal black holes in AdS and asymptotically flat spacetime. In this work, we shall see whether these proposals hold for eternal black holes in de-Sitter spacetime or not. We would like to mention that our work does not take into account certain subtleties in graviational theories, for example diffeomorphism invariance, which enables an arbitrary definition of a subregion. A discussion on this aspect can be found in \cite{Geng:2020fxl,Geng:2021hlu,Raju:2021lwh} which shows that it can have important implications to quantum gravity.
%The role that mutual information plays has been investigated in \cite{Saha:2021ohr,RoyChowdhury:2022awr}. In \cite{Saha:2021ohr}, authors have shown how the mutual information plays an important role in obtaining the time independent result of the entropy of Hawking radiation. On the other hand in our previous work \cite{RoyChowdhury:2022awr} we have given two proposals regarding the satuartion of the mutual information in two different time domains. But the above mentioned works are done by considering either eternal asymptotically flat black hole or eternal asymptotically AdS black hole. But there is no investigations on the black hole spacetime which are asymptotically de-Sitter. Further,
%on to the fine-grained entropy of Hawking radiation of BTZ black hole. It has been demonstrated that the requirement of vanishing mutual information between the subsystems leads to a time independent expression for the fine-grained entropy of the Hawking radiation that is consistent with the proper Page curve. 
\section{Brief discussion on the Kottler spacetime}
\noindent The Schwarzschild de-Sitter (SdS) spacetime metric is the unique solution of Einstein's vacuum field  equation with positive cosmological constant in $(3+1)$- spacetime dimensions. This solution is sometimes also denoted as the Kottler solution. The metric of the SdS solution has the following form \cite{https://doi.org/10.1002/andp.19183611402}
\begin{eqnarray}\label{sds}
	ds^{2}&=&-f(r)dt^{2}+\frac{dr^{2}}{f(r)}+r^{2}(d\theta^{2}+\sin^{2}\theta d\phi^{2});\nonumber\\
	f(r)&=&1-\frac{2M}{r}-\frac{\Lambda r^{2}}{3}~
\end{eqnarray}
where $M$ is the mass parameter and $\Lambda$ is the cosmological constant. The above given lapse function in terms of the AdS radius can be recast as
\begin{eqnarray}
f(r)=1-\frac{2M}{r}-\frac{r^{2}}{L^{2}_{\mathrm{AdS}}}~~.
\end{eqnarray}
We can recover the asymptotically flat Schwarzschild spacetime in the limit $\Lambda\rightarrow0$ (or $L_{\mathrm{AdS}}\rightarrow \infty$).\\
We shall now discuss about the horizon structure of the Kottler metric. One can show that the horizon structure depends on the value of the cosmological constant $(\Lambda)$ as there exists a critical value of $\Lambda=\Lambda_{crit}=\frac{1}{9M^{2}}$ above which there event horizon does not exists and the corresponding solution is then denoted as the naked singularity. However, in the range $0<\Lambda<\Lambda_{crit}$ (or $\frac{m}{L_{\mathrm{AdS}}}<\frac{1}{3\sqrt{3}}$), there are three solutions for $f(r)=0$. Out of these three solutions only two are physical solutions \cite{https://doi.org/10.1002/andp.19183611402,Gibbons:1977mu}, one is known as the black hole horizon $(r_{H})$ and the other one is known as the cosmological horizon $(r_{c})$, $r_{c}> r_{H}$. Furthermore, in the limit $\Lambda\rightarrow\Lambda_{crit}$ there is a degenerate horizon \cite{https://doi.org/10.1002/andp.19183611402}. In this work, we will only consider the range $0<\Lambda<\Lambda_{crit}$ along with the following form of the lapse function \cite{Goswami:2022ylc}
\begin{eqnarray}\label{lf}
	f(r)=\frac{1}{L_{\mathrm{AdS}}^{2}r}(r_{H}-r)(r-r_{c})(r+r_{H}+r_{c})~.
\end{eqnarray}
The expressions for the $r_{H}$ and $r_{c}$ (in terms of the mass parameter and the cosmological constant) are obtained to be \cite{Gibbons:1977mu,Bhattacharya:2013tq,Yadav:2022jib}
\begin{eqnarray}\label{ho}	r_{H}&=&\frac{2}{\sqrt{\Lambda}}\cos\left(\frac{\pi}{3}+\frac{\arccos(3M\sqrt{\Lambda})}{3}\right)\nonumber\\
r_{c}&=&\frac{2}{\sqrt{\Lambda}}\cos\left(\frac{\pi}{3}-\frac{\arccos(3M\sqrt{\Lambda})}{3}\right)~.
\end{eqnarray}
In order to proceed further, we shall now rewrite the metric in the Kruskal coordinates. As there are two different choices available for the horizons, there exists two different sets of Kruskal coordinates. This is due to the reason that the Kruskal coordinate transformations contain surface gravity in the expression which has different values corresponding to the different event horizons. This we denote as $\kappa_{H}$ (associated to the black hole horizon $r_{H}$) and $\kappa_{c}$ (associated to the cosmological horizon $r_{c}$). Keeping this mind, one can show two alternative forms of the metric in terms of the two different Kruskal coordinates. This can be represented as black horizon representation of the metric and the cosmological horizon description of the metric.\\
\noindent In order to obtain the form of the metric in the Kruskal coordinates, we first introduce the tortoise coordinate which satisfies the following transformation
\begin{eqnarray}
u&=&t-r^{*}(r)~,~v=t+r^{*}(r)
\end{eqnarray}
\begin{widetext}
where $r^{*}(r)$ is the tortoise coordinate, given as
\begin{eqnarray}
r^{*}(r)=\alpha_{H}\ln(|r_{H}-r|)-\alpha_{c}\ln(|r-r_{c}|)+\alpha^{\prime}\ln(r+r_{H}+r_{c}).
\end{eqnarray}
\end{widetext}
The expressions of $\alpha_{H}$, $\alpha_{c}$ and $\alpha^{\prime}$ read
\begin{eqnarray}
	\alpha_{H}&=&\frac{L^{2}_{\mathrm{AdS}}r_{H}}{(r_{c}-r_{H})(2r_{H}+r_{c})}\nonumber\\
	\alpha_{c}&=&\frac{L^{2}_{\mathrm{AdS}}r_{c}}{(r_{c}-r_{H})(2r_{c}+r_{H})}\nonumber\\
	\alpha^{\prime}&=&\frac{L^{2}_{\mathrm{AdS}}(r_{H}+r_{c})}{(2r_{c}+r_{H})(2r_{H}+r_{c})}~.
\end{eqnarray}
%\textbf{\textcolor{red}{Follow Yadav's paper for this section and give the relevant expressions.}}\\
\noindent We first introduce the black hole horizon description of the metric. For the right wedge of the black hole horizon, the Kruskal coordinates read %\textbf{\textcolor{red}{Delete this `around' word. I am not seeing this word either in Yadav's paper or in Narayan's paper. Use something meaningful.}}
\begin{eqnarray}
U_{H}&=&-e^{-\kappa_{H}(t-r^*(r))}\nonumber\\
V_{H}&=&e^{\kappa_{H}(t+r^*(r))}	
\end{eqnarray}
and for the left wedge it read
\begin{eqnarray}
	U_{H}&=&e^{\kappa_{H}(t+r^*(r))}\nonumber\\
	V_{H}&=&-e^{-\kappa_{H}(t-r^*(r))}	
\end{eqnarray}
where $\kappa_{H}$ is the surface gravity associated to black hole horizon
\begin{eqnarray}
	\kappa_{H}=\frac{(r_{c}-r_{H})(2r_{H}+r_{c})}{2L^{2}_{\mathrm{AdS}}r_{H}}~.
\end{eqnarray}
%\textbf{\textcolor{red}{The above expression is self-contradictory as you are saying it is the surface gravity associated to the black hole horizon, however, it has a dependency on cosmological horizon. In order to resolve this you have to write these expressions in terms of the cosmological constant, as Yadav has shown in his paper.}}
\noindent Further, one can obtain the following form of the Hawking temperature associated to the black hole horizon
 \begin{eqnarray}
 	T_{H}=\frac{\kappa_{H}}{2\pi}=\frac{(r_{c}-r_{H})(2r_{H}+r_{c})}{4\pi L^{2}_{\mathrm{AdS}}r_{H}}=\frac{1}{\beta_{H}}~.
 \end{eqnarray}
\noindent In terms of the cosmological constant and the mass parameter, the surface gravity $(\kappa_{H})$ and Hawking temperature associated to the black hole horizon read \cite{Yadav:2022jib}
\begin{widetext}
\begin{eqnarray}
	\kappa_{H}&=&\sqrt{\Lambda}\left[\frac{1}{4\cos\left(\frac{1}{3}\arccos(3M\sqrt{\Lambda})+\frac{\pi}{3}\right)}-\cos\left(\frac{1}{3}\arccos(3M\sqrt{\Lambda})+\frac{\pi}{3}\right)\right]\\
	T_{H}&=&\frac{\sqrt{\Lambda}}{2\pi}\left[\frac{1}{4\cos\left(\frac{1}{3}\arccos(3M\sqrt{\Lambda})+\frac{\pi}{3}\right)}-\cos\left(\frac{1}{3}\arccos(3M\sqrt{\Lambda})+\frac{\pi}{3}\right)\right]~.
\end{eqnarray}	
On the other hand, the Bekenstein-Hawking entropy for the black hole horizon is given by $S_{BH}=\frac{\pi r_{H}^{2}}{G_{N}}$. Finally, the black horizon description of the metric in terms of the Kruskal coordinate reads
\begin{eqnarray}\label{bhp}
	ds^{2}=-F^{2}(r)dU_{H}dV_{H}+r^{2}\Omega^{2}_{2}~;~F^{2}(r)=\frac{f(r)}{\kappa_{H}^{2}}e^{-2\kappa_{H}r^{*}(r)}\nonumber\\
\end{eqnarray}
\noindent where the detailed expression of $F(r)$ has the following form
	\begin{eqnarray}
		F(r)=\frac{2L_{\mathrm{AdS}} r_{H}}{\sqrt{r}}\frac{|r-r_{c}|^{\frac{1}{2}(1+\frac{r_{c}}{r_{H}}\left(\frac{2r_{H}+r_{c}}{2r_{c}+r_{H}}\right))}(r+r_{c}+r_{H})^{\frac{1}{2}(1-\frac{r_{c}^{2}-r_{H}^{2}}{r_{H}(2r_{c}+r_{H})})}}{(2r_{H}+r_{c})(r_{c}-r_{H})}~.
	\end{eqnarray}
\end{widetext}		
\noindent We noe move on to describe the metric in terms of the cosmological horizon. The Kruskal coordinates for the right wedge of the cosmological horizon read
\begin{eqnarray}
U_{c}&=&-e^{-\kappa_{c}(t-r^*(r))}\nonumber\\
V_{c}&=&e^{\kappa_{c}(t+r^*(r))}	
\end{eqnarray}
and for the left wedge, it read 
\begin{eqnarray}
	U_{c}&=&e^{\kappa_{c}(t+r^*(r))}\nonumber\\
V_{c}&=&-e^{-\kappa_{c}(t-r^*(r))}~.	
\end{eqnarray}
 The surface gravity $(\kappa_{c})$ and the Hawking temperature associated to the cosmological horizon have the following respective forms
\begin{eqnarray}\label{sgc}
	\kappa_{c}&=&\frac{(r_{c}-r_{H})(2r_{c}+r_{H})}{2L^{2}_{\mathrm{AdS}}r_{c}}\\
	T_{c}&=&\frac{\kappa_{c}}{2\pi}=\frac{(r_{c}-r_{H})(2r_{c}+r_{H})}{4\pi L^{2}_{\mathrm{AdS}}r_{c}}=\frac{1}{\beta_{c}}~.
\end{eqnarray}
\noindent The form of $\kappa_{c}$ and $T_{c}$ in terms of the cosmological constant and mass parameter are given as \cite{Yadav:2022jib}
\begin{widetext}
	\begin{eqnarray}
	\kappa_{c}&=&\sqrt{\Lambda}\left[\frac{1}{4\cos\left(\frac{1}{3}\arccos(3M\sqrt{\Lambda})-\frac{\pi}{3}\right)}-\cos\left(\frac{1}{3}\arccos(3M\sqrt{\Lambda})-\frac{\pi}{3}\right)\right]\\
	T_{c}&=&\frac{\sqrt{\Lambda}}{2\pi}\left[\frac{1}{4\cos\left(\frac{1}{3}\arccos(3M\sqrt{\Lambda})-\frac{\pi}{3}\right)}-\cos\left(\frac{1}{3}\arccos(3M\sqrt{\Lambda})-\frac{\pi}{3}\right)\right]~.	
	\end{eqnarray}
Therefore the cosmological horizon description of the metric in terms of the Kruskal coordinates can be written down as 
\begin{eqnarray}\label{cpm}
	ds^{2}=-G^{2}(r)dU_{c}dV_{c}+r^{2}d\Omega_{2}^{2}~;~ G^{2}(r)=\frac{f(r)}{\kappa_{c}^{2}}e^{-2\kappa_{c}r^{*}(r)}\nonumber\\
\end{eqnarray}
where the conformal factor $G(r)$ has the following form
\begin{eqnarray}
	G(r)=\frac{2L_{\mathrm{AdS}}r_{c}}{\sqrt{r}}\frac{|r_{H}-r|^{\frac{1}{2}\left(1-\frac{r_{H}}{r_{c}}\frac{r_{H}+2r_{c}}{r_{c}+2r_{H}}\right)}(r+r_{c}+r_{H})^{\frac{1}{2}(1+\frac{r_{c}^{2}-r_{H}^{2}}{r_{c}(2r_{H}+r_{c})})}}{(r_{c}-r_{H})(2r_{c}+r_{H})}~.
\end{eqnarray}
\begin{figure}[htb]
	\centering
	\includegraphics[scale=0.5]{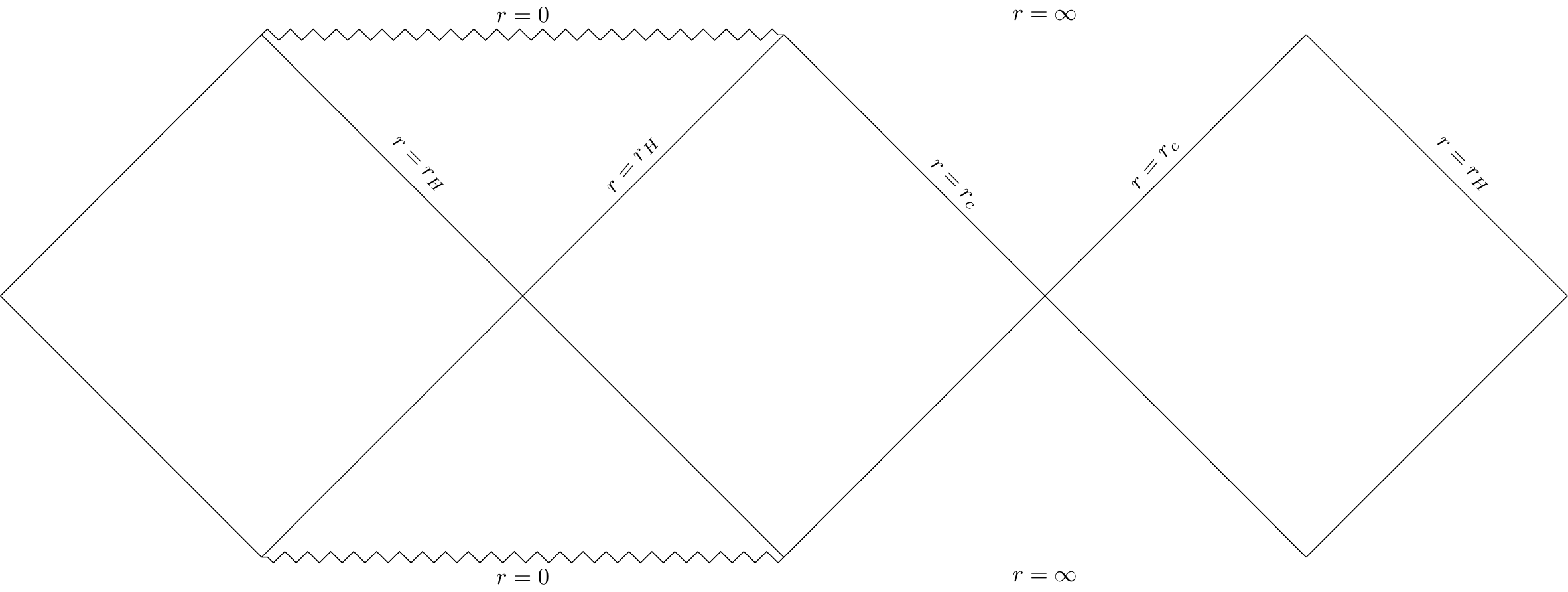}
	\caption{Penrose-Carter diagram of Schwarzschild de-Sitter spacetime. In the above, $r=r_{H}$ represents the black hole event horizon and $r=r_{c}$ is the cosmological event horizon.}
	\label{fig0}
\end{figure}
\end{widetext}
The above mentioned two alternative descriptions of the SdS metric can be understood in terms of the Penrose-Carter diagrams. This we provide in Fig.\eqref{fig0} where two physical horizons have been pointed out either of which can be used to describe the spacetime equivalently. In this work, our aim is to study the Page curve of radiation associated to both Hawking radiation and Gibbons-Hawking radiation. This can be done by isolating different patches of the spacetime by introducing the thermal opaque membrane \cite{10.1143/PTP.122.1515,ma2017thermodynamic,sekiwa2006thermodynamics,gomberoff2003sitter,PhysRevD.105.065007,PhysRevD.73.084009}. These patches have been denoted as the black hole patch and the cosmological patch in the literature. We have shown this in Fig.\eqref{figPatch}. 
\begin{figure}[!h]
	\begin{minipage}[t]{0.48\textwidth}
		\centering\includegraphics[width=\textwidth]{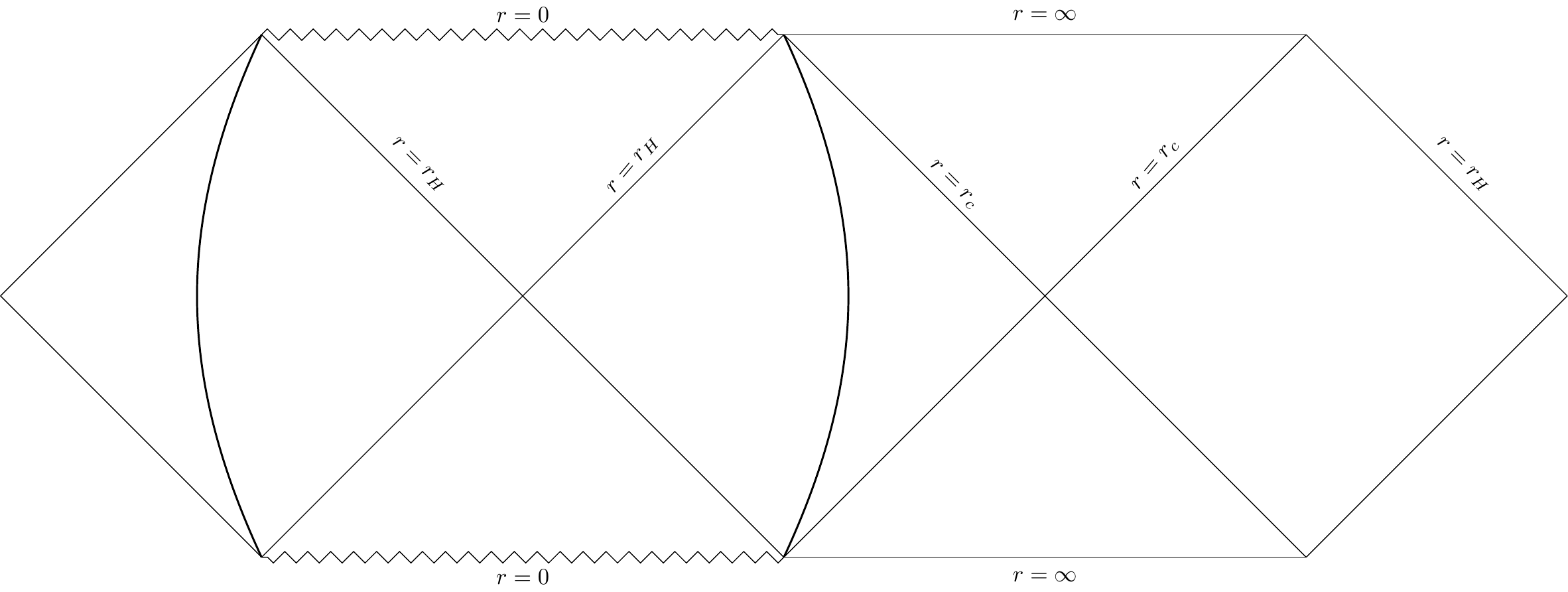}\\
		{\footnotesize  Black hole patch}
	\end{minipage}\hfill
	\begin{minipage}[t]{0.48\textwidth}
		\centering\includegraphics[width=\textwidth]{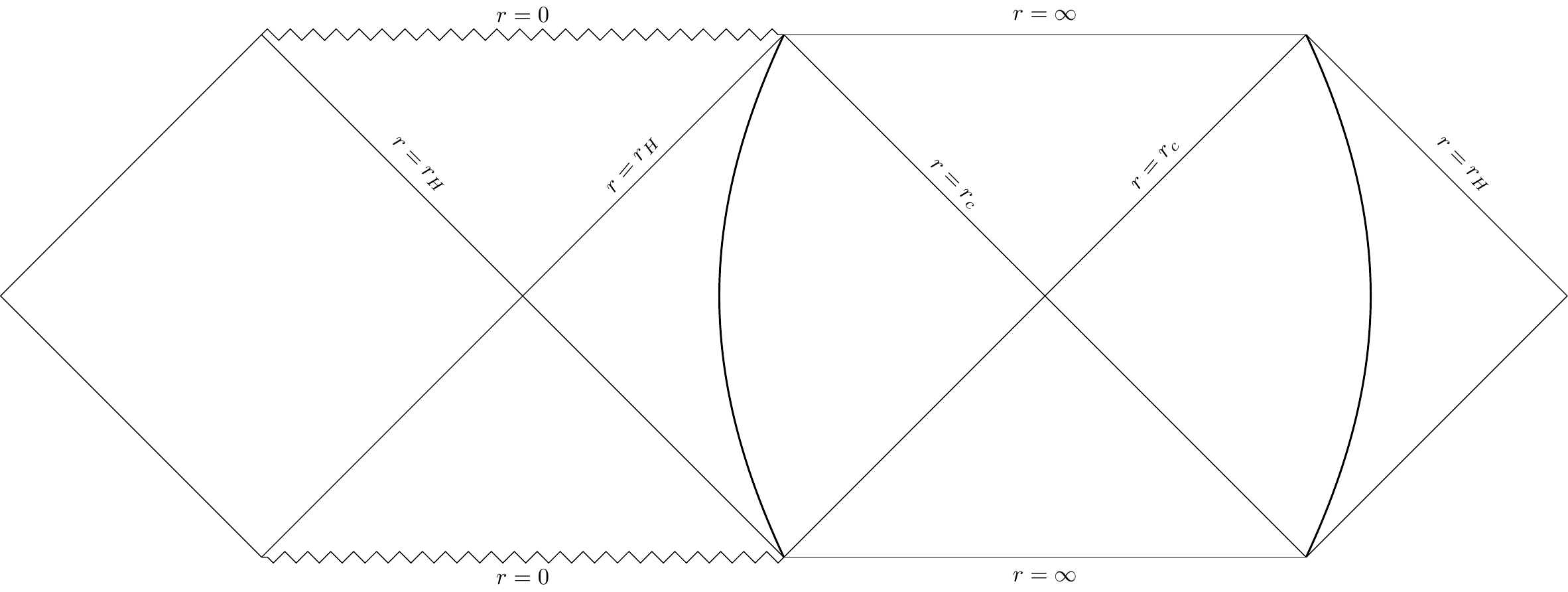}\\
		{\footnotesize Cosmological patch}
	\end{minipage}
	\caption{SAdS spacetime with thermal opaque membrane.}
	\label{figPatch}
\end{figure}
The principal reason behind introducing this thermal opaque membrane lies in the fact that we do not have any Kruskal coordinates which can remove the coordinate singularities simultaneously from both the black hole horizon and the cosmological horizon. On the other hand for the Schwarzschild de-Sitter spacetime the two horizons, namely the back hole  horizon and the cosmlogical horizon can be thought as two different thermodynamic systems with different temperatures. Therefore they are not in the thermal equilibrium. For non-equilibrium system it is very much difficult to study its thermodynamic properties. Therefore to make the analysis simpler one has to ensure that the system (either the black hole horizon or the cosmological horizon) is in thermal equilibrium. The thermal opaque membrane does this job \cite{10.1143/PTP.122.1515,ma2017thermodynamic,sekiwa2006thermodynamics,gomberoff2003sitter,PhysRevD.105.065007,PhysRevD.73.084009}. In a multi horizon spacetime one can use thermal opaque membrane to analyse one horizon by taking the other one as boundary. One can understand this thermal opaque membrane by following the approach given in \cite{Yadav:2022jib,PhysRevD.105.065007}.\\
Let us consider the radial part of the Klein-Gordon equation in SdS spacetime, which is found to be \cite{Yadav:2022jib,PhysRevD.105.065007}
\begin{eqnarray}
\left(-\frac{\partial^{2}}{\partial t^{2}}+\frac{\partial^{2}}{\partial r^{*2}}\right)\psi(r)+V_{eff}(r)\psi(r)=0~~.
\end{eqnarray} 
The explicit form of the effective potential $(V_{eff})$ can be obtained by using the lapse function given in eq.(\ref{sds}). This reads \cite{Yadav:2022jib,PhysRevD.105.065007}
\begin{eqnarray}
	V_{eff}=\left(1-\frac{2M}{r}-\frac{\Lambda r^{2}}{3}\right)\left(\frac{l(l+1)}{r^{2}}+\frac{2M}{r^{3}}-\frac{\Lambda}{3}\right)\nonumber~.
\end{eqnarray}
One can show that the above expression vanishes for both the black hole horizon and the cosmological horizon. In \cite{Yadav:2022jib,PhysRevD.105.065007} it was shown that this effective potential can be treated as the partition between the black hole and cosmological horizons. To understand this in the Penrose diagram one can introduce the Kruskal time-like and space-like coordinates for the black hole patch as
\begin{eqnarray}
U_{H}=T_{H}-R_{H}~,~V_{H}=T_{H}+R_{H}
\end{eqnarray}
and similarly for the cosmological patch
\begin{eqnarray}
U_{c}=T_{c}-R_{c}~,~V_{c}=T_{c}+R_{c}~.
\end{eqnarray}
By using this above Kruskal time-like and space-like coordinates, one can obtain the following \cite{Yadav:2022jib}
\begin{eqnarray}
	-U_{H}V_{H}&=&R_{H}^{2}-T_{H}^{2}=e^{2\kappa_{H}r^{*}(r)}\\
	-U_{c}V_{c}&=&R_{c}^{2}-T_{c}^{2}=e^{2\kappa_{c}r^{*}(r)}~.
\end{eqnarray}
The above results suggest that for $r=constant$, a hyperbola (membrane) in the $R_{H(c)}-T_{H(c)}$ plane can be realized in both the black hole and the cosmological patch.\\
\noindent On the other hand, it has been suggested that the analogue of ``defect" in wedge holography is nothing but the ``thermal opaque membrane" in Schwarzschild de-Sitter eternal black hole. Gravity may be considered to be sufficiently weak at these membranes because the membrane in question is far from the black hole/de-Sitter patch. We now proceed to investigate the role of mutual information of various subsystems in the Page curve associated to a multi-event horizon black hole spacetime.
% In this work we will study the information paradox for both the black hole patch and cosmological patch. In order to study the information paradox in the black hole patch we need to freeze the other event horizon and the vice versa. This can be done by placing a thermal opaque membrane \cite{10.1143/PTP.122.1515,ma2017thermodynamic,sekiwa2006thermodynamics,gomberoff2003sitter} in between. The analogue of ``defect" in wedge holography is the ``thermal opaque membrane" in Schwarzschild de-Sitter black holes. Gravity may be considered to be sufficiently weak at these membranes because the membrane in question is far from the black hole/de-Sitter patch, and 2D CFT formulas can be used to determine the entanglement entropy of Hawking radiation. Although our arrangement doesn't use holography, this is one way to think about it.
 %The topic of how to characterise double holography/wedge holography for "multi-event horizon black holes" arises naturally in light of this statement. This, in our opinion, will be a fantastic description to explain how black holes in multi-event horizon spacetimes solve the information dilemma.\\
\section{Analysis for the black hole patch} 
\noindent We now proceed to study the Page curve of Hawking radiation for the SdS eternal black hole in $3+1$ dimensions. As we have mentioned already, in order to probe the Hawking radiation we need to restrict ourselves to the black hole patch by introducing the thermal opaque membrane to freeze the cosmological horizon. We shall work with the form of the metric given in eq.(\ref{bhp}) which corresponds to the black hole horizon description of the SdS solution.\\
On the other hand, we assume that the whole spacetime is filled with conformal matter of central charge $c$. To be more precise, we will consider the matter to be a free CFT. We will incorporate the $s$-wave approximation in conformal matter sector \cite{Polchinski:2016hrw,Hubeny:2009rc,Hashimoto:2020cas}. The reason behind this is that the process of the Hawking radiation is dominated by the $s$-wave modes. Under this approximation we can neglect the angular part of the metric. So we can compute the entanglement entropy of the Hawking radiation by using the $2d$ CFT formula \cite{Calabrese:2009ez,Calabrese:2009qy}. Further, the $s$-wave approximation in the matter sector also implies that we can neglect the massive modes of the matter fields. We can ignore these massive modes of the matter fields because the entangling regions are very far apart from each other and therefore the theory of the conformal matter fields reduces to the $2d$ conformal field theory.\\
In this work, our motivation is to check whether the proposals given in \cite{Saha:2021ohr,RoyChowdhury:2022awr} (where the analysis is restricted only to the eternal black holes in asymptotically AdS  and flat spacetime or else) hold for a spacetime geometry with the positive cosmological constant. Particualrly, in this section we study the black hole patch of the Schwarzschild de-Sitter spacetime and check whether the results reported in \cite{Saha:2021ohr,RoyChowdhury:2022awr} holds or not. \\
As mentioned earlier, The black hole patch is equivalent to the Penrose diagram of the flat Schwarzschild black hole embedded in the de Sitter spacetime with cosmological horizons in both sides. We will focus on two scenarios here. Firstly, we will discuss what happens before the Page time $(t_{H}^{Page})$, then we will proceed to probe the after Page time scenario. In the before the Page time scenario, we intend to discuss the role of mutual information between $R^{+}_{H}$ and $R^{-}_{H}$ (shown in the Penrose diagram Fig.(\ref{fig1})) on the Page curve, as there is no island contribution in the entropy of the Hawking radiation in this time domain. However, in the after Page time scenario one has to consider the contribution from the island region which resides in the black hole interior.
 %Further, we will show that there exits a time (which is less than the Page time) at which the $I(R_{H}^{+}:R_{H}^{-})$ vanishes. This particular time is nothing  but the Hartman-Maldacena time. On the other hand, after the Page time the island starts to contribute in the entropy of Hawking radiation. In this time domain we will analyse the mutual information between $B_{H}^{+}$ and $B_{H}^{-}$. We will show that the condition of vanishing mutual information gives us time indepenent result of the von-Neumann entropy of the Hawking radiation. Further we have also obtained the expression of the scrambling time from the condition of vanishing mutual information of matter fileds on $B^{+}_{H}$ and $B^{-}_{H}$.
\subsection{Before Page time scenario: the role of $I(R^{+}_{H}:R^{-}_H)$}\label{seca}
\noindent In the scenario before the Page time scenario, that is for $t_{obs}< t_{H}^{Page}$, the entanglement entropy of the Hawking radiation can be computed by calculating the von-Neumann entropy of the matter fields on two disjoint intervals $R_{H}^{+}$ and $R_{H}^{-}$. This gives us 
$S(R_{H})=S_{vN}(R_{H}^{+}\cup R_{H}^{-})$, where $R_{H}=R_{H}^{+}\cup R_{H}^{-}$ (where the $\pm$ signifies the right and left wedges of the Pensrose-Carter diagram Fig.(\ref{fig1})).\\
%For our analysis we will take the matter filed to the $2$-d conformal matter. So in this time domain we can neglect the island contribution. Therefore the entropy of the Hawking radiation $S(R_{H})=S_{vN}(R_{H})$\\
%\noindent In this work, we first discuss the before Page time ($t_p$) scenario. For $t_{obs}<t_p$, the Hawking saddle point of the gravitational path integral dominates and the fine grained entropy of the Hawking radiation is identified by the von Neumann entropy of the matter fields.
%\begin{figure}[htb]
%	\centering
%	\includegraphics[scale=0.55]{onlyR.pdf}
	%\caption{Penrose diagram of eternal black hole of JT gravity + flat auxiliary thermal bath system. The $R_{\pm}$ regions have been shown in green with the inner boundaries $b_{\pm}=(\pm t_b,b)$ \cite{He:2021mst}.}
	%\label{fig1}
%\end{figure}
The endpoints of the disjoint regions $R^{\pm}_{H}$ are $[e^{\pm}_{H}:b^{\pm}_{H}]$. As $R^{\pm}_{H}$ regions are extended to spatial infinity (upto the thermal opaque membrane) from the inner boundary $b^{\pm}_{H}=(\pm t_{b_{H}},b_{H})$, we introduce the point $e^{\pm}_{H}$ in order to regularize it, that is, $e^{\pm}_{H}=(0,e_{H})$. We will eventually take the limit $e_H\rightarrow\infty$. In this set up, the fine-grained entropy of radiation reads
\begin{eqnarray}\label{eq5}
S_{vN}(R_{H}) = S_{vN}(R_{H}^+\cup R^{-}_H)~,~ R_{H} =S_{vN}(R_{H}^{c})~
\end{eqnarray}
where $R_{H}^{c}$ is the complement region of $R_{H}=R^{+}_H\cup R^{-}_H$. In the above, we have assumed that the state on the full Cauchy slice is a pure state. As mentioned before we consider the matter fields to be $2$d free conformal matter which can be obtained by incorporating $s$-wave approximation. 
\begin{widetext}
	So to compute the fine grained entropy of the Hawking radiation we will use the following expression 
	\begin{eqnarray}\label{sr}
		S_{vN}(R_{H}^{c})=\left(\frac{c}{3}\right)\log d(b^{+}_H,b^{-}_H)~.	
	\end{eqnarray}
\begin{figure}[h!]
	\centering
	\includegraphics[scale=.5]{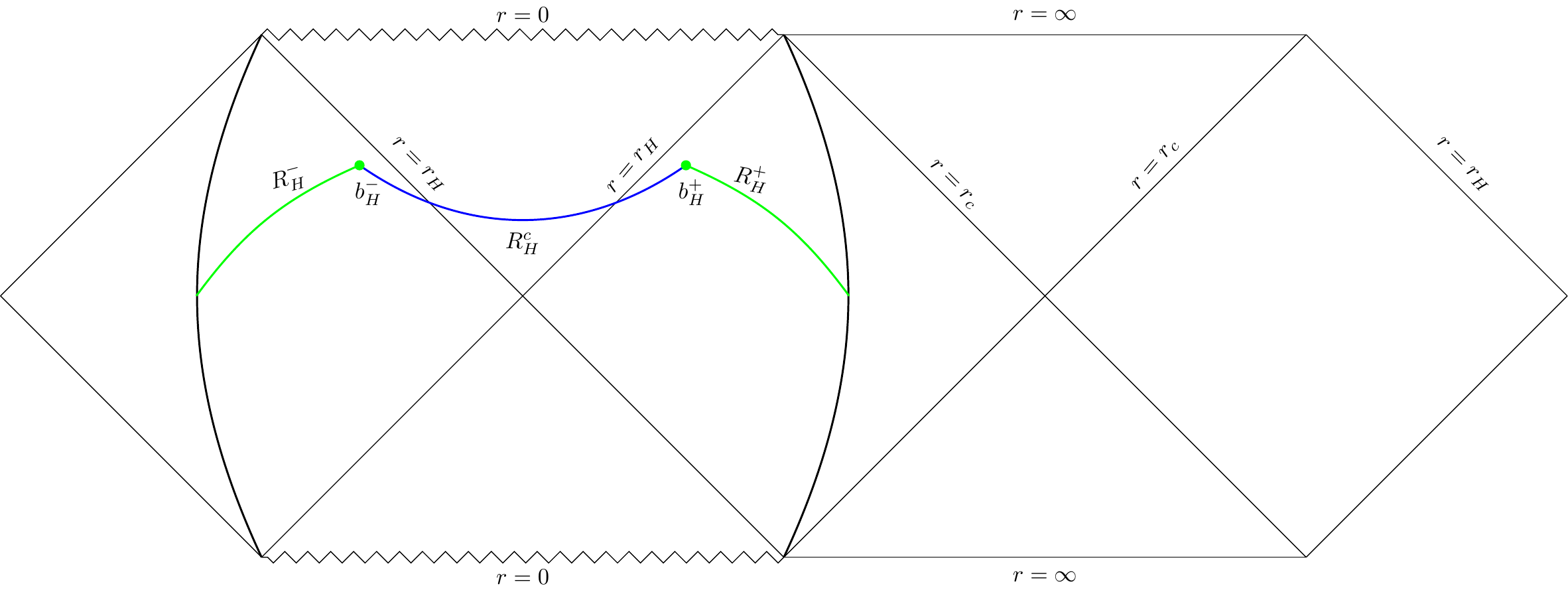}
	\caption{Penrose diagram of Schwarzschild de-Sitter black hole with thermal opaque membrane covering the cosmological patch. The $R_{H}^{\pm}$ regions are shown by green curve with $b_{H}^{\pm}=(\pm t_{b_{H}},b_{H})$. The blue line indicates the complementary region of $R_{H}=R_{H}^{+}\cup R_{H}^{-}$.}
	\label{fig1}
\end{figure}
\end{widetext}
The distance $d(b^{+}_H,b^{-}_H)$, given in the above expression can be computed explicitly from the metric given in eq.\eqref{bhp}. This reads 
%The expression of $d(b^{+}_{H},b^{-}_{H})$ is given by
\begin{eqnarray}
d(b_{H}^{+},b_{H}^{-})=2F(b_{H}) e^{\kappa_{H} r^{*}(b_{H})}\cosh(\kappa_{H} t_{b_{H}})~.
\end{eqnarray}
\begin{widetext}
\noindent Now using the above expression in eq.(\ref{sr}), the entanglement entropy of Hawking radiation is found to be
\begin{eqnarray}\label{Neweq1}
S(R_{H})&=&S_{vN}(R^{+}_H\cup R^{-}_H)\nonumber\\ &=&\left(\frac{c}{3}\right)\log\left[\left(\frac{\beta_{H}}{\pi}\right) \sqrt{f(b_{H})}\cosh\left(\frac{2\pi t_{b_{H}}}{\beta_{H}}\right)\right].	
\end{eqnarray}
\end{widetext}
From the above result we can observe that in the early time domain, that is for $t_{b_{H}}\ll\beta_{H}$, the fine grained entropy of Hawking radiation reduces to the following form
\begin{eqnarray}
S(R_{H}) &\approx& \left(\frac{c}{3}\right)\log\left[\left(\frac{\beta_{H}}{\pi}\right) \sqrt{f(b_{H})}\right]+\left(\frac{c}{6}\right)\left(\frac{2\pi t_{b_{H}}}{\beta_{H}}\right)^2~.\nonumber\\	
\end{eqnarray}
However, at late times ($t_{b_{H}}\gg\beta_{H}$), we obtain the following form of the entropy of the Haking radiation
\begin{eqnarray}
S(R_{H}) &\approx&	\left(\frac{c}{3}\right)\log\left[\left(\frac{\beta_{H}}{\pi}\right) \sqrt{f(b_{H})}\right]+\left(\frac{c}{3}\right)\left(\frac{2\pi t_{b_{H}}}{\beta_{H}}\right).\nonumber\\
\end{eqnarray}
From the above analysis we observe that as long as there is no island contribution,  the entanglement entropy of the Hawking radiation increases with respect to the observer's time. However, the nature of this time evolution of $S_{vN}(R_{H})$ is strikingly  different for these two different time doamins. To be precise, in the early time $S_{vN}(R_{H})$ shows quadratic behaviour with time, that is $S_{vN}(R_{H})\sim t_{b_{H}}^{2}$, and in the late time domain it grows linearly in time, that is $S_{vN}(R_{H})\sim t_{b_{H}}$.
This observation firmly agrees with the one shown in \cite{Hartman:2013qma}.\\
One can also compute the entanlement entropy of the matter fields localized on the individual regions $R_H^{+}$ and $R_H^{-}$. This can be written down as
\begin{eqnarray}\label{sr+}
	S_{vN}(R_{H}^{\pm})&=&\left(\frac{c}{3}\right)\log d(b_{H}^{\pm},e_{H}^{\pm})~.
\end{eqnarray}
We can compute the above given distances by using the black hole metric given in eq.(\ref{bhp}). The expressions of $d(b_{H}^{+},e_{H}^{+})$ and $d(b_{H}^{-},e_{H}^{-})$ read
\begin{widetext}
\begin{eqnarray}
	d(b_{H}^{+},e_{H}^{+})&=&\left[2F(b_{H})F(e_{H})e^{\kappa_{H} r^{*}(b_{H})}
	\left(\cosh(\kappa_{H}r^{*}(b_{H}))
	-\cosh(\kappa_{H}t_{b_{H}})\right)\right]^{\frac{1}{2}}=d(b_{H}^{-},e_{H}^{-})~.
\end{eqnarray}
\noindent In the above result we assume that $\mathop{\textrm{Limit}}_{e\rightarrow \infty}r^{*}(e)=0$. By substituting the above expression in eq.(\ref{sr+}), 
%The explicit computation of the above distances (by using the flat metric given in eq.(\ref{bath metric}) 
we get the following results
\begin{eqnarray}\label{NEq2}
S_{vN}(R^{+}_{H})=S_{vN}(R^{-}_{H})=\left(\frac{c}{6}\right)	\log\left[2\left(\frac{\beta_{H}}{2\pi}\right)^2\sqrt{f(b_{H})f(e_{H})}\left\{|\cosh\left(\frac{2\pi r^*(b_{H})}{\beta_{H}}\right)-\cosh\left(\frac{2\pi t_{b_{H}}}{\beta_{H}}\right)|\right\}\right]~~~.		
\end{eqnarray}
\noindent Now with the computed results (given in eq.\eqref{Neweq1} and eq.\eqref{NEq2}) in hand one can obtain the expression for the mutual information (MI) between the matter fields localised on the region $R^{+}_{H}$ and $R^{-}_{H}$. This is obtained to be
%\end{widetext}
% we will compute the mutual information between the regions $R_{H}^{+}$ and $R_{H}^{-}$	by using the expressions given in eq.\eqref{Neweq1} and eq.\eqref{Neweq2}. Therefore the mutual information between $R_{H}^{+}$ and $R_{H}^{-}$ reads
%\begin{widetext}
\begin{eqnarray}\label{Newqeq3}
	I(R_{H}^{+}:R_{H}^{-}) &=& S_{vN}(R_{H}^{+})+ S_{vN}(R_{H}^{-})-S_{vN}(R_{H}^{+}\cup R_{H}^{-})\nonumber\\
	&=&\left(\frac{c}{3}\right)\log\left[\left(\frac{\beta_{H}}{2\pi}\right)\sqrt{f(e_{H})}\left\{\frac{|\cosh\left(\frac{2\pi r^*(b_{H})}{\beta_{H}}\right)-\cosh\left(\frac{2\pi t_{b_{H}}}{\beta_{H}}\right)|}{\cosh\left(\frac{2\pi t_{b_{H}}}{\beta_{H}}\right)}\right\}\right]~.
\end{eqnarray}
In order to understand the behaviour of MI thoroughly (for both early and late time scenario), we compute its form by considering the justified limits. In the early time domain $(t_{b_{H}} \ll \beta_{H})$, the expression of mutual information reduces to the following form
%\end{widetext}
\begin{eqnarray}\label{Newqeq4}
	I(R_{H}^{+}:R_{H}^{-})&\approx&\left(\frac{c}{3}\right)\left[\log\left[\left(\frac{\beta_{H}}{2\pi}\right)\sqrt{f(e_{H})}\cosh\left(\frac{2\pi r^*(b_{H})}{\beta_{H}}\right)\right]-\mathrm{sech}\left(\frac{2\pi r^*(b_{H})}{\beta_{H}}\right)-\left(\frac{2\pi^2}{\beta_{H}^2}\right)\left\{1+\mathrm{sech}\left(\frac{2\pi r^*(b_{H})}{\beta_{H}}\right)\right\}t_{b_{H}}^2\right]~.\nonumber\\
\end{eqnarray}
The above expression suggests that at the early time domain $I(R_{H}^{+}:R_{H}^{-})$ decreases with the time-scaling $\sim t_{b_{H}}^{2}$. On the other hand, at the late times ($t_{b_{H}} \gg \beta_{H}$), we obtain the following form of the mutual information
	\begin{eqnarray}\label{Newqeq}
I(R_{H}^{+}:R_{H}^{-})&\approx&\left(\frac{c}{3}\right)\left[\log\left[\left(\frac{\beta_{H}}{2\pi}\right)\sqrt{f(e_{H})}\right]-2\cosh\left(\frac{2\pi r^*(b_{H})}{\beta_{H}}\right)e^{-\left(\frac{2\pi t_{b_{H}}}{\beta_{H}}\right)}\right]~.\nonumber\\		
	\end{eqnarray}
\end{widetext}
This in turn means that at late times ($t_{b_{H}} \gg \beta_{H}$), $I(R_{H}^{+}:R_{H}^{-})$ increases with respect to the observer's time $t_{b_{H}}$. Interestingly, one can note by looking at eq.(s)\eqref{Newqeq4} and \eqref{Newqeq} that there exists a particular value of $t_{b_{H}}$ at which the mutual information will be zero and the entanglement wedge corresponding to $R_{H}^{+}\cup R_{H}^{-}$ will be in its disconnected phase\footnote{As we know mutual information between two subsystems, namely, $A$ and $B$ satisfies the non-negative property, that is, $I(A:B)\geq0$. This means zero is the lowest possible value mutual information can have where the correlation between $A$ and $B$ vanishes}. This observation supports the following proposal given in \cite{RoyChowdhury:2022awr}
\begin{widetext}
% This implies that there is a time-scale (we denote as $t_b=t_R$), at which the above mentioned transition occurs. We intend to evaluate this time $t_R$ by the following proposal.
	\noindent\textbf{Proposal I:} \textit{ For an eternal black hole in de-Sitter spacetime, starting from a finite, non-zero value (at $t_{b_{H}}=0$), the mutual information between $R_{H}^{+}$ and $R_{H}^{+}$ vanishes at  a particular value of the observer's time ($t_{b_{H}}=t_H$).}\\
	
%\begin{widetext}
%\noindent We therefore need to compute the von Neumann entropy of two disjoint intervals of two dimensional conformal field theory. This is given by \cite{Calabrese:2009ez}	
%\begin{eqnarray}\label{eq6}
%S_{vN}(R_+\cup R_-)= \left(\frac{c}{3}\right)\log\Big[\frac{d(\Lambda_+,b_+)d(b_+,b_-)d(b_-,\Lambda_-)d(\Lambda_-,\Lambda_+)}{d(\Lambda_+,b_-)d(\Lambda_-,b_+)}\Big]~.
%\end{eqnarray}
%\end{widetext} 
%{One can observe that the term $\left(\frac{c}{3}\right)\log d(\Lambda_+,\Lambda_-)$ corresponds to von Neumann entropy of the state on the full Cauchy slice. As we have mentioned earlier, the state on the full slice is a pure state so the contribution of this term is zero.} 
\noindent Now we will compute the expression of the time scale $t_{H}$ at which the mutual information between $R_{H}^{+}$ and $R_{H}^{-}$ vanishes. In 
order to do this we will use the expression given in eq.(\ref{Newqeq3}) along with the above given proposal. This reads
\begin{eqnarray}\label{eq7}
	I(R_{H}^{+}:R_{H}^{-})|_{t_{b_{H}}=t_H}&=&0~.
\end{eqnarray}
\noindent One can solve the above equation to obtain the value of $t_H$. This is found to be
%\begin{widetext}
\begin{eqnarray}
	t_H=\left(\frac{\beta_{H}}{2\pi}\right)\cosh^{-1}\left\{\left(\frac{\frac{\beta_{H}}{2\pi}\sqrt{f(e_{H})}}{1+\frac{\beta_{H}}{2\pi}\sqrt{f(e_{H})}}\right)\cosh\left(\frac{2\pi r^*(b_{H})}{\beta_{H}}\right)\right\}
	.\nonumber\\
\end{eqnarray}
\noindent The above expression suggests that the time scale $t_{H}$ is much smaller than $t_{b_{H}}=\beta_{H}$, that is $t_{H}\ll\beta_{H}$. Therefore the time scale $t_{H}$ lies in the early time domain.
%{It is to be noted that $t_R$ is much smaller in amplitude in comparison with the time scale $t_b = \beta$ or in other other words, this particular time-scale $t_b=t_R$ resides in the early time domain as $t_R\ll\beta$.}
The expression of $S_{vN}(R_{H}^{+}\cup R_{H}^{-})$ at this particular time ($t_{b_{H}}=t_H$) reads
%In obtaining the above, we have used lapse function of eternal black hole of JT gravity. Now if we compute the exact value of $S_{vN}(R_+\cup R_-)$ at this particular time, that is $t_b=t_R$, we obtain
\begin{eqnarray}\label{eq12}
S_{vN}^{t_{b_{H}}=t_{H}}(R_{H}^{+}\cup R_{H}^{-})&=&\frac{c}{3}\log\left[\frac{\left(\frac{\beta_{H}\sqrt{f(e_{H})}}{2\pi}\right)^{2}}{1+\frac{\beta_{H}\sqrt{f(e_{H})}}{2\pi}}\cosh\left(\frac{2\pi r^{*}(b_{H})}{\beta_{H}}\right)\right]\nonumber\\
&\approx&\frac{c}{3}\log\left[\frac{\beta_{H}}{2\pi}\sqrt{f(e_{H})}\right]+\frac{c}{6}\left(\frac{r_{H}}{b_{H}}\right)^{2}~.\nonumber\\	
\end{eqnarray}
\end{widetext}
%{In the last line we have use the fact that $\frac{\beta\sqrt{f(\xi)}}{2\pi}\gg1$ and $b\gg r_{+}$.\\}
Our proposal suggests that the mutual correlation between $R_{H}^{+}$ and $R_{H}^{-}$ is non-zero for the time interval $0\leq t_{b_{H}}<t_H$.
The value of $I(R_+:R_-)$ is maximum at $t_{b_{H}}=0$ and then it decreases for the range $t_{b_{H}}\leq t_H$, and vanishes exactly at $t_{b_{H}}=t_H$. Further, it also depicts the fact that the associated entanglement wedge of $R_{H}^{+}\cup R_{H}^{-}$ is in connected phase initially. Then, at $t_{b_{H}}=t_H$, the mutual information between $R_{H}^{+}$ and $R_{H}^{-}$ vanishes and the entanglement wedge associated to $R_{H}^{+}\cup R_{H}^{-}$ makes the transition to the disconnected phase. Once again we would like to mention that $t_H\ll \beta_{H}$. These observations strongly indicate that this time $t_H$
\pagebreak
 is nothing but the Hartman-Maldacena time $t_{HM}$, as reported in our previous work \cite{RoyChowdhury:2022awr}. Furthermore, the expression of mutual information, $I(R_{H}^{+}:R_{H}^{-})$ at $t_{b_{H}}=\beta_{H}$ is obtained to be
\begin{widetext}
\begin{eqnarray}
I(R_{H}^{+}:R_{H}^{-}) &=&\left(\frac{c}{3}\right)\log\left[\left(\frac{\beta_{H}}{2\pi}\right)\sqrt{f(e_{H})}\cosh\left(\frac{2\pi r^*(b_{H})}{\beta_{H}}\right)\right]~.\nonumber\\
\end{eqnarray}
\end{widetext}
The above result tells us that after the Hartman- Maldacena time,  the mutual correlation between $R_{H}^{+}$ and $R_{H}^{-}$ ($I(R_+:R_-)$) starts to increase with respect to the observer's time $t_{b_{H}}$. 
\subsection{After Page time scenario: probing the role of $I(B_{H}^{+}:B_{H}^{-})$}\label{sec2}
\noindent We now proceed to discuss the after Page time scenario $t_{b_{H}}\ge t_{H}^{Page}$. Just after the Page time $t_{H}^{Page}$, the island starts to contribute. This in turn means that one has to generalize the concept of entanglement entropy by introducing the concept of fine-grained entropy. This generalization incorporates the area term in the formula given in eq.(\ref{eq1}) along with the island contribution.
 %So after Page time the fine-grained entropy of radiation is obtained from eq.(\ref{eq1})
One can observe that the term $S_{vN}(I_{H}\cup R_{H})$ satisfies the identity $S_{vN}(I_{H}\cup R_{H}^{+}\cup R_{H}^{-})=S_{vN}(B_{H}^{+}\cup B_{H}^{-})$. The regions of $B_{H}^{\pm}$ can be specified as $(b_{H}^{\pm} \rightarrow a_{H}^{\pm})$ where $a_{H}^{\pm}=(\pm t_{a_{H}},a_{H})$ are the end points of the island. This can be understood by the Penrose diagram, given in Fig(\ref{fig2}). Now as we have mentioned earlier, in this work we are considering $2d$ free CFT as the matter sector. This in turn means that the expression associated to $S_{vN}(B_{H}^{+} \cup B_{H}^{-})$ can be evaluated by using the following formula \cite{Calabrese:2009ez}
%\begin{figure}[htb]
%	\centering
%	\includegraphics[scale=0.55]{withIsland.pdf}
	%\caption{Penrose diagram specifying the island region (in red) with boundaries $a_{\pm}=(\pm t_a,a)$ and the $B_{\pm}$ regions (in blue). The inner boundaries of $B_{\pm}$ regions are $a_{\pm}=(\pm t_a,a)$ and the outer boundaries are $b_{\pm}=(\pm t_b,b)$ \cite{He:2021mst}.}
	%\label{fig2}
%\end{figure}
%where the value of $I(R_+:R_-)$ is maximum at $t_b=0$ and decreases upto, starting from a non-zero finite value (at $t_b=0$), the upto a particular value ($t_b=t_R$) of the observer's time $t_b$ (that is $t_b<t_R$), the mutual correlation between $R_+$ and $R_-$ is non-zero and the associated entanglement wedge of $R_+\cup R_-$ is in connected phase. This also means that one can denote the time-scale $t_b<t_R$ as the early time domain. However, at the time $t_b=t_R$, the mutual information between $R_+$ and $R_-$ vanishes and the associated entanglement wedge of $R_+\cup R_-$ becomes disconnected \cite{Takayanagi:2017knl}. Keeping this in mind, one can denote the time-scale $t_b> t_R$ as the late-time domain. Furthermore, in the domain $t_b<t_R$, the fine-grained entropy of Hawking radiation scales as $S(R)\sim t_b^2$ and for the domain $t_b> t_R$, it scales as $S(R)\sim t_b$. This can be checked by making an expansion of the expression of $S(R)=S_{vN}(R_+\cup R_-)$ \cite{Saha:2021ohr}.\\ 
\begin{widetext}
\begin{eqnarray}\label{eq13}
S_{vN}(B_{H}^{+}\cup B_{H}^{-})=\left(\frac{c}{3}\right)\log\Big[\frac{d(a_{H}^{+},a_{H}^{-})d(b_{H}^{+},b_{H}^{+})d(a_{H}^{+},b_{H}^{+})d(a_{H}^{-},b_{H}^{-})}{d(a_{H}^{+},b_{H}^{-})d(a_{H}^{-},b_{H}^{+})}\Big]~.
\end{eqnarray}
\begin{figure}[h!]
	\centering
	\includegraphics[scale=0.5]{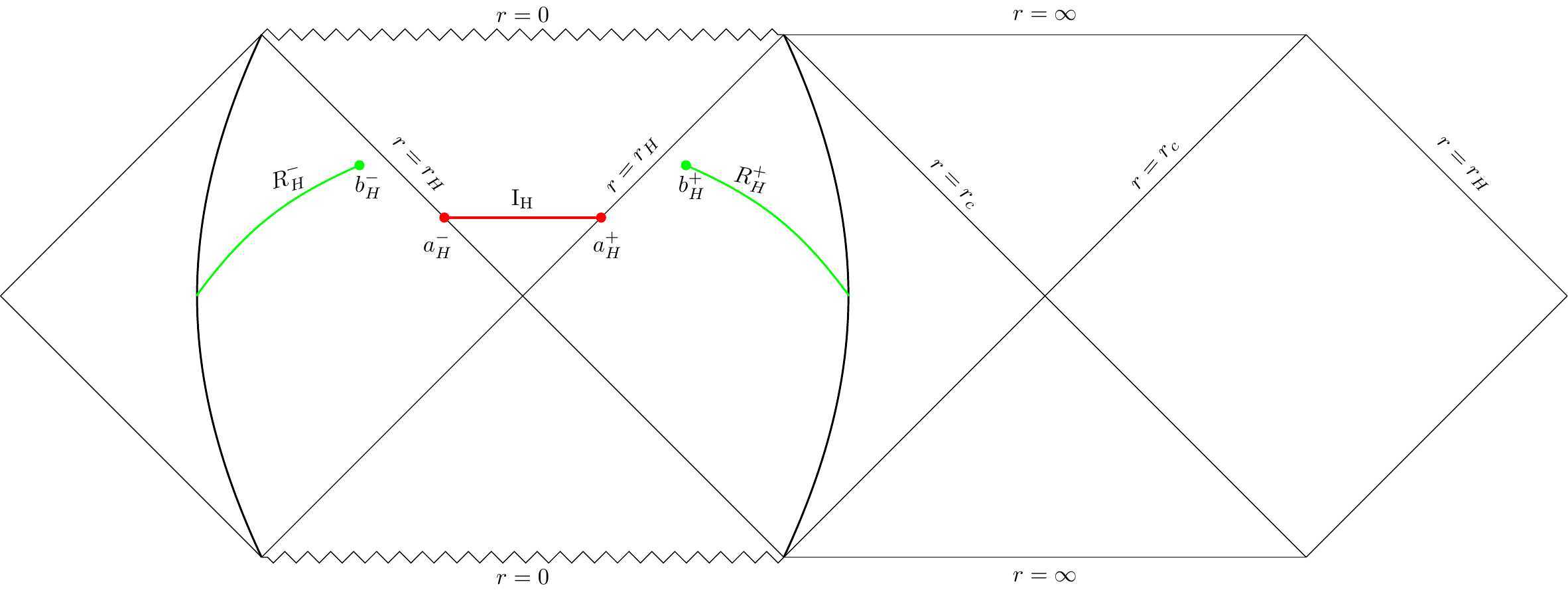}
	\caption{Penrose diagram of the black hole patch (with thermal opaque membrane covering the cosmological patch) indicating the island region (in red) with endpoints $a_{H}^{\pm}=(\pm t_{a_{H}},a_{H})$. The radiation regions are shown by the green line.}
	\label{fig2}
\end{figure}
\end{widetext}
Now, in order to compute the explicit form of the entanglement entropy of the matter fields, we have to substitute the distances in eq.\eqref{eq13}. This can be calculated from the black hole metric given in eq.\eqref{bhp}. 
%{One can show that for the black hole patch gravity eternal black hole solution, one obtains the following}
%\begin{eqnarray}\label{eq14}
%	S_{vN}(B_{H}^+\cup B_{H}^-)&=&S_{vN}(B_{H}^{+})+S_{vN}(B_{H}^{-})\nonumber\\
%	&&+\sim\mathcal{O}(e^{-\frac{2\pi t_{a_{H}}}{\beta_{H}}}) +\sim\mathcal{O}(e^{-\frac{2\pi t_{b}}{\beta}})
%\end{eqnarray}
% where $S_{vN}(B_{\pm})=\left(\frac{c}{3}\right)\log d(b_{\pm},a_{\pm})$. 
In recent works in this direction, it has been suggested that at the late times ($t_{a_{H}},t_{b_{H}}\gg\beta_{H}$), one can make the following approximation \cite{Hashimoto:2020cas,Matsuo:2020ypv}
\begin{eqnarray}\label{eq15}
S_{vN}(B_{H}^{+}\cup B_{H}^{-}) \approx S_{vN}(B_{H}^{+})+ S_{vN} (B_{H}^{-})	
\end{eqnarray}
where
\begin{eqnarray}\label{sb+}
S_{vN}(B_{H}^{\pm})=\left(\frac{c}{3}\right)\log d(b_{H}^{\pm},a_{H}^{\pm})
\end{eqnarray}
 By using the above mentioned approximation in eq.\eqref{eq1} along with the correct \textit{area term} and upon extremization one can show that the final expression for $S(R_{H})$ is nothing but $S(R_{H})$, that is $S(R_{H})=2S_{BH}+...$~. This has already been shown in \cite{Yadav:2022jib,Krishnan:2020fer}. In \cite{Saha:2021ohr,RoyChowdhury:2022awr} it was shown that the approximation given in eq.(\ref{eq15}) corresponds to the fact that one has to ignore the terms $\sim e^{-\frac{2\pi t_{b_{H}}}{\beta_{H}}}$. This in turn means that the approximation $e^{-\frac{2\pi t_{b_{H}}}{\beta_{H}}} \approx 0$ is associated with the vanishing of mutual information $(I(B_{H}^{+}:B_{H}^{-})=S_{vN}(B_{H}^{+})+ S_{vN} (B_{H}^{-})- S_{vN}(B_{H}^{+}\cup B_{H}^{-})\approx0)$, only at the leading order. However, if the contribution from the terms $\sim e^{-\frac{2\pi t_{b_{H}}}{\beta_{H}}}$ are kept, then it will eventually give us a time-dependent expression of $S(R)$. This issue was addressed in our previous works \cite{Saha:2021ohr,RoyChowdhury:2022awr}. We now extend our previous study for the de-Sitter spacetime by proposing the following
  %It is to be noted that the above late-time approximation (given in eq.\eqref{eq15}) indirectly gives the hint of vanishing mutual correlation (only in the leading order) between $B_{H}^+$ and $B_{H}^-$. Keeping this in mind, we give our \textit{after Page time proposal} \cite{Saha:2021ohr,RoyChowdhury:2022awr}. This reads
\begin{widetext}
\noindent\textbf{Proposal II:}  \textit{For an eternal black hole in de-Sitter spacetime, the mutual information between the black hole subsystems $B_{H}^+$ and $B_{H}^-$ vanishes just after the Page time when the island starts to contribute.} 	
\end{widetext}	
Holographically the above proposal implies that just after the Page time, when the replica wormhole saddle points starts to dominate, the entanglement wedge of $B_{H}^{+}\cup B_{H}^{-}$ makes the transition from connected to disconnected phase \cite{Takayanagi:2017knl,Saha:2021kwq,Chowdhury:2021idy} and this results in $I(B_{H}^{+}:B_{H}^{-})=0$. Now, according to the above given proposal, we need to compute the following
\begin{eqnarray}\label{eq16}
	I(B_{H}^{+}:B_{H}^{-})&=&0\nonumber\\
S_{vN}(B_{H}^{+})+S_{vN}(B_{H}^{-})&=&S_{vN}(B_{H}^{+}\cup B_{H}^{-})~.	
\end{eqnarray} 
\noindent By substituting the explicit expressions from eq.(\ref{sb+}) and eq.(\ref{eq13}), one obtains the following equality
%\begin{widetext}
\begin{eqnarray}\label{neweq4}
d(a_{H}^{+},b_{H}^{-})(a_{H}^{-},b_{H}^{+})&=&d(a_{H}^{+},a_{H}^{-})d(b_{H}^{+},b_{H}^{-})~.
\end{eqnarray}
%\end{widetext}
Substituting this equality in $S_{vN}(B^{+}_{H}\cup R^{-}_{H})$ (given in eq.(\ref{eq13})), we obtain
\begin{eqnarray}\label{neweq5}
S_{vN}(B_{H}^{+}\cup B_{H}^{-})=\frac{c}{3}\log\left(d(a_{H}^{+},b_{H}^{+})d(a_{H}^{-},b_{H}^{-})\right) ~.
\end{eqnarray}
By using the  metric of the black hole patch given in eq.\eqref{bhp}, one can compute the explicit expressions corresponding to the mentioned various distances.
\begin{widetext}
This reads 
\begin{eqnarray}
d(a_{H}^{\pm},b_{H}^{\pm})&=&\sqrt{2F(a_{H})F(b_{H})e^{\kappa_{H}(r^{*}(b_{H})+r^{*}(a_{H}))}}\Big[\cosh[\kappa_{H}(r^{*}(a_{H})-r^{*}(b_{H}))]-\cosh[\kappa_{H}(t_{a_{H}}-t_{b_{H}})]\Big]^{\frac{1}{2}}\label{neweq6}\\
d(a_{H}^{\pm},b_{H}^{\mp})&=&\sqrt{2F(a_{H})F(b_{H})e^{\kappa_{H}(r^{*}(b_{H})+r^{*}(a_{H}))}}\Big[\cosh[\kappa_{H}(r^{*}(a_{H})-r^{*}(b_{H}))]+\cosh[\kappa(t_{a_{H}}+t_{b_{H}})]\Big]^{\frac{1}{2}}\label{neweq7}\\
d(b_{H}^{+},b_{H}^{-})&=&2F(b_{H})e^{\kappa_{H} r^{*}(b_{H})}\cosh(\kappa_{H} t_{b_{H}})\label{neweq8}\\
d(a_{H}^{+},a_{H}^{-})&=&2F(a_{H})e^{\kappa_{H} r^{*}(a_{H})}\cosh(\kappa_{H} t_{a_{H}})\label{neweq9}~.
\end{eqnarray}
\end{widetext}
These above expressions of distances suggests that
\begin{eqnarray}\label{dd}
d(a^{+}_{H},b^{+}_{H})&=&d(a^{-}_{H},b^{-}_{H})\nonumber\\
d(a^{+}_{H},b^{-}_{H})&=&d(a^{-}_{H},b^{+}_{H}).
\end{eqnarray}
 This in turn means that we can recast the expression of $S_{vN}(B^{+}_{H}\cup B^{-}_{H})$ (given in eq.(\ref{neweq5})) in the following form
\begin{eqnarray}\label{sbb+}
	S_{vN}(B_{H}^{+}\cup B_{H}^{-})=\frac{2c}{3}\log d(a_{H}^{+},b_{H}^{+})
\end{eqnarray}
On the other hand, substituting these expressions of distances in eq.\eqref{neweq4} along with fact given in eq.(\ref{dd}), we obtain the following condition
\begin{eqnarray}\label{eq17}
t_{a_{H}}-t_{b_{H}}=|r^*(a_{H})-r^*(b_{H})|~.
\end{eqnarray}
The above obtained condition is very interesting as it enables us express $t_{a_{H}}$ in terms of the other quantities. By using this mentioned property in eq.\eqref{sbb+}, we obatin the entanglement entropy of the conformal matter fields
\begin{eqnarray}\label{eq18}
S_{vN}(B_{H}^{+}\cup B_{H}^{-})=\frac{c}{3}\log\left(\frac{2}{\kappa_{H}^{2}}\right)+\frac{c}{6}\log[f(a_{H})f(b_{H}))]~.	\nonumber\\
\end{eqnarray}
The importance of the above result lies in the fact that it is independent of time. Now if we substitute the above expression in eq.\eqref{eq1} together with the \textit{area term}, that is $\frac{\mathrm{Area}(\partial I_{H})}{4G_N}=2\times\frac{4\pi a_{H}^{2}}{4G_N}$, the fine grained entropy of the Hawking radiation reads
\begin{eqnarray}\label{fgb}
	S(R_{H})=2\times\frac{4\pi a_{H}^{2}}{4G_N}+\frac{c}{3}\log\left(\frac{2}{\kappa_{H}^{2}}\right)+\frac{c}{6}\log[f(a_{H})f(b_{H})]~.\nonumber\\
\end{eqnarray}
We now need to find the value of the island parameter `$a_{H}$'. This we obtain by performing the extremization of the above result. This leads to the following value
\begin{eqnarray}\label{eq19}
	a_{H}=r_{H} - \left(\frac{cG_N }{24\pi}\right)\frac{1}{r_H}+...~.
\end{eqnarray}
The above results shows that the quantum extremal surfaces are located inside the black hole event horizon \cite{Yadav:2022jib,Goswami:2022ylc}. However, in case of eternal black holes in AdS, it has been noted that the quantum extremal surfaces reside just outside the event horizon \cite{Saha:2021ohr,RoyChowdhury:2022awr}. So, the position of the island endpoints are different for dS and AdS spacetime. Substitution of the above extremized value of $``a_{H}"$ in eq.(\ref{fgb}) leads to the following expression of fine-grained entropy of Hawking radiation
\begin{eqnarray}\label{eq20}
	S(R_{H})=2S_{BH} +\frac{c}{3}\log\left(S_{BH}\right)-\frac{\left(\frac{c}{2}\right)^2}{2S_{BH}}+...\nonumber\\
\end{eqnarray}
It can be noted from the above expression that it is time independent and contains logarithmic and inverse power law correction terms \cite{Saha:2021ohr, RoyChowdhury:2022awr}. Revisiting the condition $I(B_+:B_-)=0$ (given in eq.\eqref{eq17}) with the obtained value of $``a_{H}"$ (given in eq.\eqref{eq20}), we get
\begin{eqnarray}\label{eq21}
	t_{a_{H}}-t_{b_{H}}= \left(\frac{\beta_{H}}{8\pi}\right)\log\left(S_{BH}\right)=t_{H}^{Scr}~
\end{eqnarray}
where $t_{H}^{Scr}$ is the \textit{Scrambling time}\cite{Sekino:2008he,Hayden:2007cs} for the black hole patch. The remarkable observation made above in turn tells that just after the Page time $t_{H}^{P}$, the replica wormhole saddle points start to dominate and the emergence of island in the black hole interior leads to the disconnected phase of the entanglement wedge $B_{H}^{+}\cup B_{H}^{-}$, characterized by the condition given in eq.\eqref{eq21}. On the other hand, the explicit expression of the Page time is found to be
\begin{eqnarray}\label{eq22}
	t_{H}^{Page} = \left(\frac{3\beta_{H}}{\pi c}\right) S_{BH}- \left(\frac{\beta_{H}}{\pi}\right)\log\left(S_{BH}\right)...~.\nonumber\\
\end{eqnarray}
In the above expression, the leading piece is the familiar form of the Page time, where the rest represent the sub-leading corrections to it.
\section{Analysis for the cosmological patch}
\noindent In this part, we will study the Page curve corresponding to the entanglement entropy of Gibbons-Hawking radiation. This we do by restricting ourselves in the cosmological patch and treating the black holes on each side as frozen (for the same reason as in the previous section, we again add two thermal opaque membranes on either side of the black hole patch). For the area of interest, the corresponding metric is given in eq.(\ref{cpm}). It has been noted that studies in this direction are often restricted to the black holes in asymptotically flat or AdS spacetimes, however the most recent data shows that the universe is expanding faster with a de-Sitter like characteristics. In this context, the cosmological event horizons emit and absorb radiation similar to the black hole event horizon and this radiation has been denoted as the Gibbons-Hawking radiation. In general, the entropy creation of the cosmic horizon is an observer-dependent feature in contrast to the black hole. It develops as a result of ignorance regarding what exists beyond the cosmological horizon.\\
Now, in order to study the cosmological patch we have to freeze the black hole patch by the thermal opaque membranes on the both sides. Once again, we will discuss two scenarios here, namely, the before cosmological Page time $(t_{c}^{Page})$ scenario and the after cosmological Page time scenario. 
%Similar to black hole patch, for cosmolodical patch we also give two propsals regarding the saturation of mutual information (or regarding the connectedness of the associated entanglement wedge). Our first proposal probes the before ``cosmological Page time" scenario. In this time domain there is no ``cosmological island" contribution of the entanglement entropy of Gibbons-Hawking radiation. Therefore in this scenario we will discuss the saturation of mutual information between $R^{+}_{c}$ and $R^{-}_{c}$. We will show that there exist a time scale which is smaller than the ``cosmological Page time" where the mutual correlation between $R^{+}_{c}$ and $R^{-}_{c}$ vanishes, and this time scale is identified with the Hartman-Maldacena time for the cosmological patch. On the other hand after $t_{c}^{Page}$, cosmological island starts to contribute in the entropy of Gibbons-Hawking radiation. In this time domain we have studied the important role played by the mutual information between $B^{+}_{c}$ and $B^{-}_{c}$. We will show that condition of vanishing mutual information between $B^{+}_{c}$ and $B^{-}_{c}$ gives us the time independent result of the entanglement entropy of Gibbons-Hawking radiation. 
\subsection{Before cosmological Page time scenario: The role of $I(R_{c}^{+}:R_{c}^{-})$} 
\noindent Similar to the previous scenario, this time domain corresponds to the facts that the observer's time is less then the cosmological Page time, that is $t_{c}\ll t_{c}^{Page}$. As mentioned earlier, in this time domain there is no cosmological island contribution. Therefore the entanglement entropy of the Gibbons-Hawking (GH) radiation ($S(R_{c})$) is given by the von Neumann entropy of the matter fields on $R_{c}=R_{c}^{+}\cup R_{c}^{-}$, that is $S(R_{c})=S_{vN}(R_{c})$. It is to be noted that the end points of the  disjoint regions $R_{c}^{\pm}$ are $[e_{c}^{\pm}:b_{c}^{\pm}]$.
\begin{widetext}
\noindent As $R^{\pm}_{c}$ regions are extended to spatial infinity (upto the thermal opaque) from the inner boundary $b^{\pm}_{c}=(\pm t_{b_{c}},b_{c})$, we introduce the point $e^{\pm}_{c}$ in order to regularize it, that is, $e^{\pm}_{c}=(0,e_{c})$. This can be visualised in the Penrose diagram given in Fig.(\ref{fig3}). We will eventually take the limit $e_c\rightarrow\infty$. Now, we need to compute the following in order to obtain the desired result
\begin{figure}[htb]
	\centering
	\includegraphics[scale=0.5]{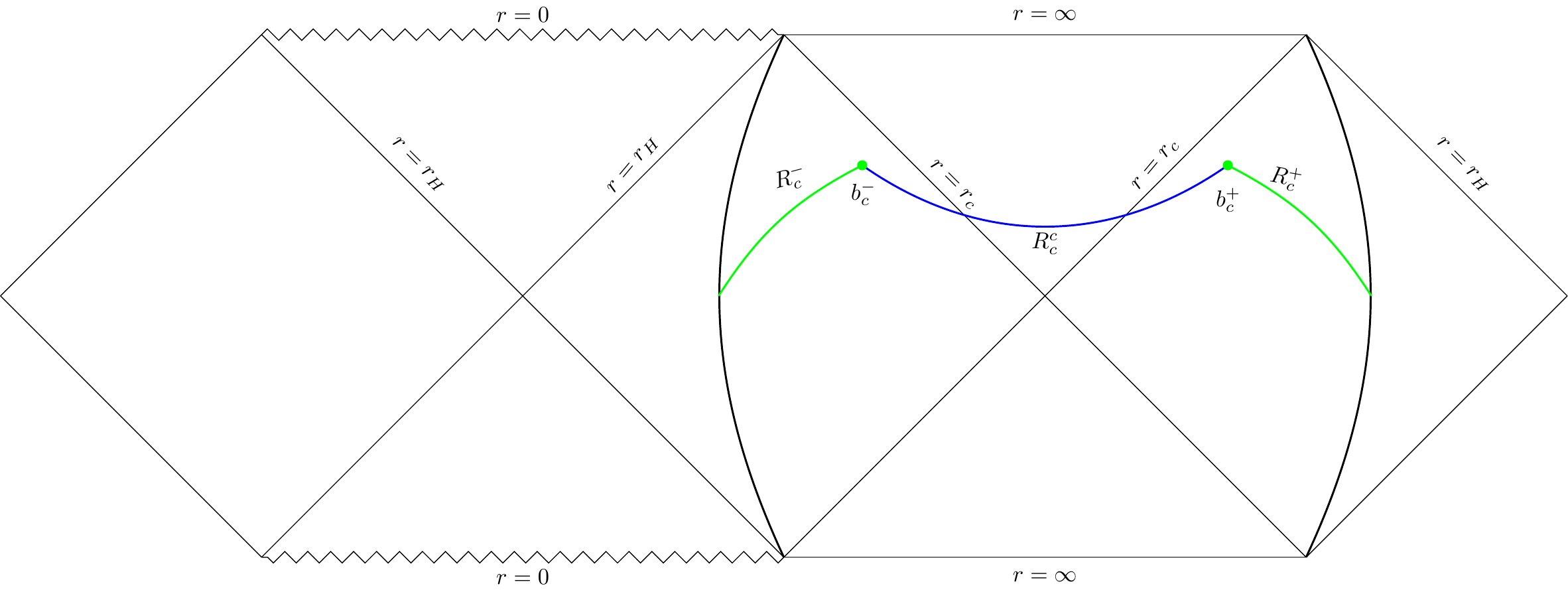}
	\caption{Penrose diagram of cosmological patch of SdS spacetime with thermal opaque. The regions $R_{c}^{\pm}$ is indicated by the green line. The complementary region of $R_{c}=R_{c}^{+}\cup R_{c}^{-}$, that is, $R_{c}^{c}$ is shown by the in figure by blue line.}
	\label{fig3}
\end{figure}
\end{widetext}
\begin{eqnarray}
S_{vN}(R_{c})&=&S_{vN}(R_{c}^+\cup R_{c}^-)~;~ R_{c} = R^{+}_{c}\cup R^{-}_{c}~.	
\end{eqnarray}
Once again we assume that the state on the full Cauchy slice is a pure state, therefore the entanglement entropy of the GH radiation reads
\begin{eqnarray}
	S_{vN}(R_{c}^+\cup R_{c}^-)&=&S_{vN}(R_{c}^{c})
\end{eqnarray}
Now, keeping in mind the $s$-wave approximation in the matter sector, we use the $2d$ conformal field theory formula. This reads
\begin{eqnarray}\label{scr}
	S_{vN}(R_{c}^{c})=\left(\frac{c}{3}\right)\log d(b^{+}_c,b^{-}_c)~.	
\end{eqnarray}
To compute the distance $d(b^{+}_c,b^{-}_c)$, in the cosmological patch we will use the metric given in eq.(\ref{cpm}). The in turn gives us
\begin{eqnarray}
d(b_{c}^{+},b_{c}^{-})=2G(b_{c}) e^{\kappa_{c} r^{*}(b_{c})}\cosh(\kappa_{c} t_{b_{c}})
\end{eqnarray}
where $\kappa_{c}$ (surface gravity of the cosmological patch) is given in eq.(\ref{sgc}). The above given result and eq.(\ref{scr}) lead us to the following result for entanglement entropy of the GH radiation 
\begin{eqnarray}\label{NewEq1}
	S(R_{c})&=&S_{vN}(R^{+}_c\cup R^{-}_c)\nonumber\\
	&=&\left(\frac{c}{3}\right)\log\left[\left(\frac{\beta_{c}}{\pi}\right) \sqrt{f(b_{c})}\cosh\left(\frac{2\pi t_{b_{c}}}{\beta_{c}}\right)\right].
\end{eqnarray}
We now follow the footstep shown in the previous section and compute the form of $S(R_{c})$ for both early and late time domains. In the early time domain ($t_{b_{c}}\ll\beta_{c}$), $S(R_{c})$ reduces to the following form
\begin{eqnarray}
	S(R_{c}) &\approx& \left(\frac{c}{3}\right)\log\left[\left(\frac{\beta_{c}}{\pi}\right) \sqrt{f(b_{c})}\right]+\left(\frac{c}{6}\right)\left(\frac{2\pi t_{b_{c}}}{\beta_{c}}\right)^2~.\nonumber\\
\end{eqnarray}
On the other hand, in late time domain ($t_{c}^{Page}>t_{b_{c}}\gg \beta_c$), it reads
\begin{eqnarray}
	S(R_{c}) &\approx&	\left(\frac{c}{3}\right)\log\left[\left(\frac{\beta_{c}}{\pi}\right) \sqrt{f(b_{c})}\right]+\left(\frac{c}{3}\right)\left(\frac{2\pi t_{b_{c}}}{\beta_{c}}\right).\nonumber\\
\end{eqnarray}
Once again we note that, in absence of the island contribution, $S(R_{c})$ exhibits quadratic behaviour over time in the early time domain $S(R_{c})\sim t_{b_{c}}^{2}$ and linearly with time for the late time domain $S_{vN}(R_{c})\sim t_{b_{c}}$.
Further, the entanglement entropy of the matter fields localized on the individual areas $R_{c}^+$ and $R_{c}^-$ are obtained to be
%\begin{widetext}
\begin{equation}\label{sc+}
	S_{vN}(R^{\pm}_{c})=\frac{c}{3}\log d(b^{\pm}_{c},e^{\pm}_{c})
\end{equation}
%\end{widetext}
\noindent Using the metric on the cosmological patch provided in eq.(\ref{cpm}), we can calculate the distances. The expressions of $d(b_{c}^+,e_{c}^+) $ and $d(b_{c}^-,e_{c}^-) $ read
\begin{widetext}
	\begin{eqnarray}
		d(b_{c}^{+},e_{c}^{+})&=&\left[2G(b_{c})G(e_{c})e^{\kappa_{c} r^{*}(b_{c})}
		\left(\cosh(\kappa_{c}r^{*}(b_{c}))
		-\cosh(\kappa_{c}t_{b_{c}})\right)\right]^{\frac{1}{2}}
		=d(b_{c}^{-},e_{c}^{-})~.
	\end{eqnarray}
In the above expression we are using the fact that, in the limit $e_{c}\rightarrow \infty$, $r^{*}(e)$ vanishes.
By replacing the aforementioned formula in the eq.(\ref{sc+}), we get the following results
	\begin{eqnarray}\label{NewEq2}
		S_{vN}(R_{c}^{+}) = S_{vN}(R_{c}^{-}) =\left(\frac{c}{6}\right)	\log\left[2\left(\frac{\beta_{c}}{2\pi}\right)^2\sqrt{f(b_{c})f(e_{c})}\left\{|\cosh\left(\frac{2\pi r^*(b_{c})}{\beta_{c}}\right)-\cosh\left(\frac{2\pi t_{b_{c}}}{\beta_{c}}\right)|\right\}\right]~.	
	\end{eqnarray}
Now, by using the expressions provided in eq.(\ref{NewEq1}) and eq.(\ref{NewEq2}), we once again compute the mutual information between $R_{c}^+$ and $R_{c}^-$. This reads
	\begin{eqnarray}\label{NewEq3}
		I(R_{c}^{+}:R_{c}^{-}) &=& S_{vN}(R_{c}^{+})+ S_{vN}(R_{c}^{-})-S_{vN}(R_{c}^{+}\cup R_{c}^{-})\nonumber\\
		&=&\left(\frac{c}{3}\right)\log\left[\left(\frac{\beta_{c}}{2\pi}\right)\sqrt{f(e_{c})}\left\{\frac{|\cosh\left(\frac{2\pi r^*(b_{c})}{\beta_{c}}\right)-\cosh\left(\frac{2\pi t_{b_{c}}}{\beta_{c}}\right)|}{\cosh\left(\frac{2\pi t_{b_{c}}}{\beta_{c}}\right)}\right\}\right]~.\nonumber\\
	\end{eqnarray}
\end{widetext}
Similar to the black hole patch analysis, one can show that in the early time domain $I(R_{c}^{+}:R_{c}^{-})$ decreases with the time-scaling $\sim t_{b_{c}}^{2}$ and for the late time domain  ($t_{b_{c}} \gg \beta_{c}$), $I(R_{c}^{+}:R_{c}^{-})$ increases with respect to the observer's time $t_{b_{c}}$. This once again points out the fact that there exits a time $t_{c}$, at which mutual information between $R_{c}^{+}$ and $R_{c}^{-}$ vanishes and the entanglement wedge associated to $R_{c}^{+}\cup R_{c}^{-}$ gets disconnected. Keeping this in mind, it can be said that the following proposal is valid also for the cosmological patch
\begin{widetext}
\noindent\textbf{Proposal I:} \textit{For an eternal black hole in de-Sitter spacetime, starting from a finite, non-zero value (at $t_{b_{c}}=0$), the mutual information between $R_{c}^{+}$ and $R_{c}^{+}$ vanishes at a particular value of the observer's time ($t_{b_{c}}=t_c$).}\\

\noindent In this case, the value of the time-scale $t_c$ is obtained to be
	\begin{eqnarray}
		t_c=\left(\frac{\beta_{c}}{2\pi}\right)\cosh^{-1}\left\{\left(\frac{\frac{\beta_{c}}{2\pi}\sqrt{f(e_{c})}}{1+\frac{\beta_{c}}{2\pi}\sqrt{f(e_{c})}}\right)\cosh\left(\frac{2\pi r^*(b_{c})}{\beta_{c}}\right)\right\}
		.\nonumber\\
	\end{eqnarray}
Once again we note that the time scale $t_c$ is substantially lower than $t_{b_{c}}=\beta_{c}$, that is $t_{c}\ll\beta_{c}$. As a result, the time scale $t_c$ belongs to the early time domain. Furthermore, the expression of $S_{vN}(R_{c}^{+}\cup R_{c}^{-})$ at this particular time reads
	\begin{eqnarray}\label{Eq12}
		S_{vN}^{t_{b_{c}}=t_{c}}(R_{c}^{+}\cup R_{c}^{-})&=&\frac{c}{3}\log\left[\frac{\left(\frac{\beta_{c}\sqrt{f(e_{c})}}{2\pi}\right)^{2}}{1+\frac{\beta_{c}\sqrt{f(e_{c})}}{2\pi}}\cosh\left(\frac{2\pi r^{*}(b_{c})}{\beta_{c}}\right)\right]\nonumber\\
		&\approx&\frac{c}{3}\log\left[\frac{\beta_{c}}{2\pi}\sqrt{f(e_{c})}\right]+\frac{c}{6}\left(\frac{r_{c}}{b_{c}}\right)^{2}~.	
	\end{eqnarray}
\end{widetext}
This in turn means that for the cosmological patch also the mutual correlation between $R_{c}^{+}$ and $R_{c}^{-}$ is non-zero for the time period $0\leq t_{b_{c}}<t_c$ and it reaches its maximum value at $t_{b_{c}}=0$. After that $I(R_{c}^{+}:R_{c}^{-})$ decreases for the range $t_{b_{c}}\leq t_{c}$ and finally disappears at $t_{b_{c}}=t_{c}$. This also reflects the fact that the connected phase of the corresponding entanglement wedge of $R_{c}^{+}\cup R_{c}^{-}$ gets disconnected at $t_{b_{c}}=t_{c}$. These findings once again clearly suggest that $t_{c}$ is the Hartman-Maldacena time for the cosmological patch. After this time (Hartman-Maldacena time) the mutual information between $R_{c}^{+}$ and $R_{c}^{-}$ increases with respect to the observer's time.
\begin{widetext}
\subsection{After Cosmological Page time scenario: The role of $I(B_{c}^{+}:B_{c}^{-})$}
\noindent Now, once again we proceed to probe the after cosmological Page time scenario. As we have mentioned earlier, Just after the cosmological Page time ($t_{c}^{Page}$) the island starts contribute to the fine grained entropy of Gibbons-Hawking radiation. 
\begin{figure}[htb]
	\centering
	\includegraphics[scale=0.5]{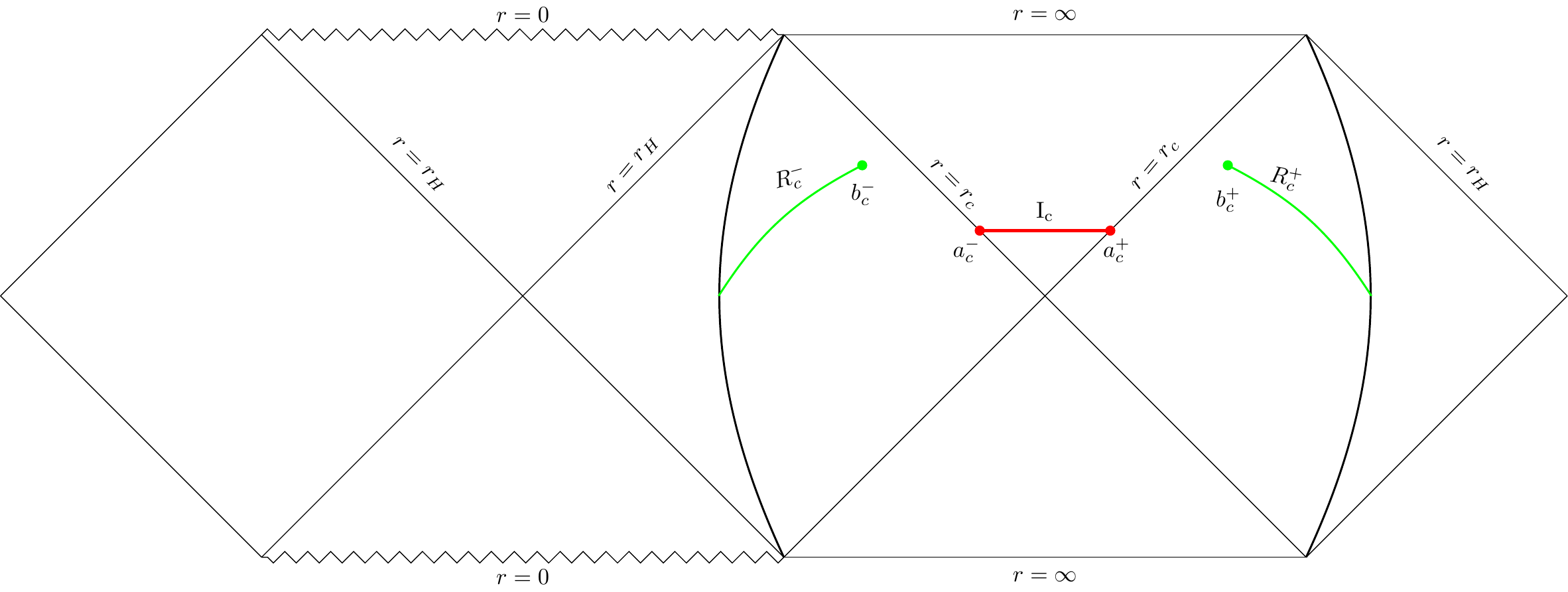}
	\caption{ The above Penrose diagram shows the cosmological patch with the thermal opaque membrane. The red line indicates the cosmological island surface with end points $a_{c}^{\pm}=[\pm t_{a_{c}},a_{c}]$. The green lines indicate the radiation regions.}
	\label{fig4}
\end{figure}
\end{widetext}
Using the fact that the matter part of eq.(\ref{eq1}) satisfies the property $S_{vN}(I_{c}\cup R_{c}^{+}\cup R_{c}^{-})=S_{vN}(B_{c}^{+}\cup B_{c}^{-})$. The regions of $B_{c}^{\pm}$ can be specified as $(b_{c}^{\pm} \rightarrow a_{c}^{\pm})$, where the island end points are pointed out as $a_{c}^{\pm}=(\pm t_{a_{c}},a_{c})$. The Penrose diagram given in Fig.(\ref{fig4}) helps us to visualise this. We now follow the steps shown in the black hole patch scenario. As we have already stated, the matter sector in this work is the $2d$ free CFT.
\begin{widetext}
As a result, the expression of $S_{vN}(B_{c}^{+}\cup B_{c}^{-})$ can be evaluated using the following formula
 \cite{Calabrese:2009ez}
 	\begin{eqnarray}\label{Eq13}
 		S_{vN}(B_{c}^{+}\cup B_{c}^{-})=\left(\frac{c}{3}\right)\log\Big[\frac{d(a_{c}^{+},a_{c}^{-})d(b_{c}^{+},b_{c}^{-})d(a_{c}^{+},b_{c}^{+})d(a_{c}^{-},b_{c}^{-})}{d(a_{c}^{+},b_{c}^{-})d(a_{c}^{-},b_{c}^{+})}\Big]~\nonumber\\
 	\end{eqnarray}
 \end{widetext}
\begin{widetext}
\noindent The distances that may be derived from the  metric provided in eq.(\ref{cpm}) must be replaced in in the above expression in order to obtain the explicit form of the entanglement entropy of the matter field. The entropy of the matter field on the individual regions can be computed by the following expression
\begin{eqnarray}
S_{vN}(B_{c}^{\pm})&=&\left(\frac{c}{3}\right)\log d(b_{c}^{\pm},a_{c}^{\pm})~.
\end{eqnarray}
Now, as we have already mentioned for the black hole patch analysis, one can compute $S_{vN}(B_{c}^{+}\cup B_{c}^{-})$ for late time by using the following approximation
\begin{eqnarray}\label{late}
	S_{vN}(B_{c}^{+}\cup B_{c}^{-})
\sim S_{vN}(B_{c}^{+})+ S_{vN}(B_{c}^{-})~.
\end{eqnarray}
\end{widetext}
The above mentioned approximation is once again associated to the fact that one has to neglect the terms $\sim e^{-\frac{2\pi t_{b_{c}}}{\beta_{c}}}$ which provides the indication of vanishing mutual correlation (only in the leading order) between $B_{c}^{+}$ and $B_{c}^{-}$. This observation is similar to the one we have already noted for the black hole patch scenario which in turn means that the following proposal should also hold for the cosmological patch 
\begin{widetext}
	\noindent\textbf{Proposal II:}  \textit{For an eternal black hole in de-Sitter spacetime, the mutual information between the between the matter fields localised on $B_{c}^+$ and $B_{c}^-$ vanishes just after the cosmological Page time.} 	
\end{widetext}
By following the same procedure we have already shown for the black hole patch, one can obtain the time-independent form of fine-grained entropy of GH radiation by using the above given proposal. This reads
\begin{eqnarray}
	S(R_{c})=2\times\frac{4\pi a_{c}^{2}}{4G_N}+\frac{c}{3}\log\left(\frac{2}{\kappa_{c}^{2}}\right)+\frac{c}{6}\log[f(a_{c})f(b_{c})]~.\nonumber\\
\end{eqnarray}
Extremising the above result with respect to the cosmological island parameter ``$a_{c}$", we get 
\begin{eqnarray}\label{ac}
		a_{c}=r_{c} - \left(\frac{cG_N }{24\pi}\right)\frac{1}{r_c}+...~.
\end{eqnarray}
The above result indicates that the cosmological island end points (quantum extremal surfaces) are located inside the cosmological horizon \cite{Yadav:2022jib}. By using the result given in eq.(\ref{ac}), we obtain the desired result of fine grained entropy of Gibbons-Hawking radiation 
\begin{eqnarray}
	S(R_{c})=2S_{GH} +\frac{c}{3}\log\left(S_{GH}\right)-\frac{\left(\frac{c}{2}\right)^2}{2S_{GH}}+...~.
\end{eqnarray}
Furthermore, the extremized value of the cosmological island parameter simplifies condition of vanishing mutual information to the following form 
\begin{eqnarray}
	t_{a_{c}}-t_{b_{c}}= \left(\frac{\beta_{c}}{8\pi}\right)\log\left(S_{GH}\right)=t_{c}^{Scr}~.
\end{eqnarray}
where $t_{c}^{Scr}$ scrambling time for the cosmological patch. Further the expressionof the cosmological Page time is obtained to be \cite{Yadav:2022jib}
\begin{eqnarray}
	t_{c}^{P} \approx \left(\frac{3\beta_{c}}{\pi c}\right) S_{GH}~.
\end{eqnarray}
\section{Conclusions}
\noindent We now provide a summary of our findings. In this work, we have tried to check whether our previously reported proposals \cite{Saha:2021ohr,RoyChowdhury:2022awr} holds for eternal black holes in de-Sitter spacetime or not. The said proposals were originally given for eternal black holes in AdS spacetime and in this work we have observed that the mentioned proposals also holds for eternal black holes in de-Sitter spacetime. The motivation to consider an eternal black hole solution in de-Sitter spacetime is associated to the subtle structure of the event-horizon for this spacetime.
\noindent We have briefly investigated the role of mutual information of various subsystems in the Page curve for both Hawking radiation and Gibbons-Hawking radiation, by keeping in mind the recent developments of the island formulation. In order to study the Page curve of above mentioned two different radiations, we have introduced the notion of thermal opaque membrane. This membrane allows us to study the two different radiations individually as it divides the whole system into two patches (equivalent descriptions), namely, the black hole patch and the cosmological patch. Further, the findings from the study of mutual information have motivated us to give two proposals for both the Black hole patch and the cosmological patch of the Schwarzschild de-Sitter spacetime.\\
The first proposal deals with the time domain where the observer's time is less than the Page time. First, we will discuss the importance of this proposal for the black hole patch. 
In this time domain the entanglement entropy of Hawking radiation does not include the island contribution. The entropy of the radiation is identified as the von-Neumann entropy of the conformal matter fields on $R_{H}^{+}\cup R_{H}^{-}$. We have incorporated the formula of $2d$ CFT in order to calculate the mentioned von Neumann entropy $S_{vN}(R_{H})= S_{vN}(R_{H}^{+}\cup R_{H}^{-})$ as we have stated that we are only considering the $s$-wave contribution of the conformal matter. In the early time domain, that is for $t_{b_{H}}\ll \beta_{H}$, we note that $S(R_{H})$ shows quadratic growth ($S(R_{H})\sim t_{b_{H}}^{2}$) and in the late time domain ($t_{b_{H}}\gg \beta_{H}$) $S(R_{H})$ increases linearly with respect to the observer's time ($S(R_{H})\sim t_{b_{H}}$). The mutual information $I(R_{H}^{+}: R_{H}^{-})$ between $R_{H}^{+}$ and $R_{H}^{-}$ is then computed by obtaining the explicit expressions of $S_{vN}(R_{H}^{+})$ and $S_{vN}(R_{H}^{-})$. With the general expression of $I(R_{H}^{+}: R_{H}^{-})$ in hand, we then proceed to investigate its behaviour in both the early and late time domain. In the early time domain, starting from the maximum value at $t_{b_{H}}=0$, $I(R_{H}^{+}:R_{H}^{-})$ starts decreasing with the time-scaling $\sim t_{b_{H}}^{2}$ and in the late time domain we find that $I(R_{H}^{+}:R_{H}^{-})$ increases with respect to $t_{b_{H}}$. This kind behaviour of the mutual information motivates us to give our first proposal which tells us that there exits a time, $t_{b_{H}}=t_{H}$ ($0<t_{H}<\beta_{H}$) at which the mutual correlation between $R_{H}^{+}$ and $R_{H}^{-}$ disappears. This in turn implies that the associated entanglement wedge $R_{H}^{+}\cup R_{H}^{-}$ becomes disconnected. Further, at $t_{b_{H}}=t_{H}$, the entropy of Hawking radiation is proportional to the logarithm of the inverse temperature of the black hole, that is $S(R_{H})|_{t_{b_{H}}=t_{H}}\sim \log\beta_{H}$. These observations indicates that this particular time-scale $t_H$ is nothing but the Hartman-Maldacena time for the black hole patch. After $t_{b_{H}}=t_{H}$, $I(R_{H}^{+}:R_{H}^{-})$ starts to increase, which in turns means that the associated entanglement wedge is once again in its connected phase. In the case of cosmological patch also we have observed similar kind of phenomena before the cosmological ``Page time" $t^{Page}_{c}$ and the Hartman-Maldacena time for the cosmological patch is denoted as $t_{c}$. The explicit expressions corresponding to both $t_H$ and $t_c$ has also been computed.\\ 
Now we will discuss about our second proposal. This proposal is associated to the time domain where the observer's time is greater than the Page time. In case of the black hole patch, after the Page time $(t_{H}^{Page})$, the entropy of Hawking radiation includes the island contribution. This inclusion of island contribution provides appropriate Page curve which portrays the time evolution of the entropy of the Hawking radiation. Following the works in this direction, it has been noted that to obtain the correct Page curve we have to use the late time approximation $S_{vN}(B_{H}^{+}\cup B_{H}^{-}) \approx S_{vN}(B_{H}^{+})+ S_{vN} (B_{H}^{-})$ \cite{Hashimoto:2020cas} which can also be understood as $I(B_{H}^{+}:B_{H}^{-})$ but only at the leading order. This approximation is associated to the fact that one has to ignore terms $\sim e^{-\frac{2\pi t_{b_{H}}}{\beta_{H}}}$. This creates a dilemma as the the core issue in this context is regarding time-dependency. However, if these terms are incorporated one gets a time-dependent form of $S(R)$ in the after Page time scenario. We address this crucial issue by demanding that the inclusion of island (replica wormhole saddle-point contributions) leads to the disconnected phase of the entanglement wedge associated to $B_{H}^{+}\cup B_{H}^{-}$. This in turns means that just after the Page time $(t_{H}^{P})$, island in turn gifts us the vanishing mutual information between $B_{H}^{+}$ and $B_{H}^{-}$. This condition of vanishing mutual information, that is $I(B_{H}^{+}:B_{H}^{-})=0$ leads to the remarkable result $t_{a_{H}}-t_{b_{H}}=t_{H}^{Scr}$ where $t_{H}^{Scr}$ is the \textit{scrambling time} \cite{Sekino:2008he,Hayden:2007cs}.  Using the subadditivity condition of von Neumann entropy we can reforge our observation in the following way. The entanglement wedge associated to $B_{H}^{+}\cup B_{H}^{+}$ is in connected phase as long as $t_{a_{H}}-t_{b_{H}}< t_{H}^{Scr}$, and when this time difference equals the scrambling time $t_{H}^{Scr}$, the entanglement wedge associated to $B_{H}^{+}\cup B_{H}^{+}$ jumps to the disconnected phase. Most importantly this condition of vanishing mutual information condition gives us the time-independent expression of the entropy of the Hawking radiation. Our proposals and observations related to mutual information gives strong realization of the concept given in \cite{Grimaldi:2022suv,VanRaamsdonk:2010pw}. For the cosmological patch also our second proposal implies that after the cosmological Page time when the island statrs contributes the entanglement wedge associated to $B_{c}^{+}\cup B_{c}^{-}$ is in the disconnected phase. Our proposal also implies that for the cosmological patch we have $t_{a_{c}}-t_{b_{c}}=t_{c}^{Scr}$, with $t_{c}^{Scr}$ is the \textit{Scrambling time} for the cosmological patch. Similar to the black hole patch scenario, we also obtain a time independent result of the entropy of the Gibbons-Haking entropy by imposing the condition of vanishing mutual information between $B_{c}^{+}$ and $B_{c}^{-}$. Another interesting fact to point out is that in both of the cases the quantum extremal surfaces lie inside the respective horizons. This behaviour is opposite to the one we observe for the eternal black hole in AdS spacetime.

\section{Acknowledgements}
\noindent ARC would like to thank SNBNCBS for Senior Research Fellowship. AS would like to acknowledge the support by Council of Scientific and Industrial Research (CSIR, Govt. of India) for the Senior Research Fellowship. The
authors would like to thank the anonymous referee for very
useful comments. 
\bibliography{Reference}
\end{document}